\documentclass[manuscript]{acmart}

\AtBeginDocument{%
  }
\usepackage{graphicx}
\usepackage{algorithmic}
\usepackage[ruled,linesnumbered]{algorithm2e}
\usepackage{fontawesome}
\usepackage{tcolorbox}
\usepackage{pifont}
\usepackage{xspace}
\usepackage{textcomp}
\usepackage{tablefootnote}
\usepackage{booktabs,multirow,stfloats}

\newcommand{\approach}{{\sc CITYWALK}\xspace}
\newcommand{\testpilot}{{\sc TestPilot}\xspace}
\definecolor{light-blue}{RGB}{0,112,192}
\definecolor{light-blue-1}{RGB}{132,193,255}
\definecolor{light-green}{RGB}{0,176,80}

\setcopyright{acmlicensed}
\copyrightyear{2024}
\acmYear{2024}
\acmDOI{XXXXXXX.XXXXXXX}

\acmJournal{TOSEM}
\acmVolume{37}
\acmNumber{4}
\acmArticle{111}
\acmMonth{8}

\begin{document}

\title[Enhancing LLM-Based C++ Unit Test Generation via \approach]{\approach: Enhancing LLM-Based C++ Unit Test Generation via Project-Dependency Awareness and Language-Specific Knowledge}

\author{Yuwei Zhang}
\email{zhangyuwei@iscas.ac.cn}
\orcid{0009-0008-1016-7361}
\author{Qingyuan Lu}
\email{luqingyuan22@otcaix.iscas.ac.cn}
\orcid{0009-0000-8024-1143}
\affiliation{%
  \institution{Institute of Software, Chinese Academy of Sciences; University of Chinese Academy of Sciences}
  \city{Beijing}
  \country{China}
}

\author{Kai Liu}
\email{kliu@sse.com.cn}
\orcid{}
\affiliation{%
  \institution{Shanghai Stock Exchange Technology Co., Ltd.}
  \city{Shanghai}
  \country{China}
}

\author{Wensheng Dou}
\authornote{Affiliated with Nanjing Institute of Software Technology, University of Chinese Academy of Sciences, Nanjing, China.}
\email{wsdou@otcaix.iscas.ac.cn}
\orcid{0000-0002-3323-0449}
\author{Jiaxin Zhu}
\authornotemark[1]
\authornote{Corresponding authors}
\email{zhujiaxin@otcaix.iscas.ac.cn}
\orcid{0000-0002-0905-2355}
\affiliation{%
  \institution{Institute of Software, Chinese Academy of Sciences; University of Chinese Academy of Sciences}
  \city{Beijing}
  \country{China}
}

\author{Li Qian}
\email{lqian@sse.com.cn}
\orcid{}
\author{Chunxi Zhang}
\email{chxzhang@sse.com.cn}
\orcid{}
\author{Zheng Lin}
\email{zhenglin@sse.com.cn}
\orcid{}
\affiliation{%
  \institution{Shanghai Stock Exchange Technology Co., Ltd.}
  \city{Shanghai}
  \country{China}
}

\author{Jun Wei}
\authornotemark[1]
\authornotemark[2]
\email{wj@otcaix.iscas.ac.cn}
\orcid{0000-0002-8561-2481}
\affiliation{%
  \institution{Institute of Software, Chinese Academy of Sciences; University of Chinese Academy of Sciences}
  \city{Beijing}
  \country{China}
}

\renewcommand{\shortauthors}{Zhang et al.}

\begin{abstract}
Unit testing plays a pivotal role in the software development lifecycle, as it ensures code quality. However, writing high-quality unit tests remains a time-consuming task for developers in practice. More recently, the application of large language models (LLMs) in automated unit test generation has demonstrated promising results. Existing approaches primarily focus on interpreted programming languages (e.g., Java), while mature solutions tailored to compiled programming languages like C++ are yet to be explored. The intricate language features of C++, such as pointers, templates, and virtual functions, pose particular challenges for LLMs in generating both executable and high-coverage unit tests. To tackle the aforementioned problems, this paper introduces \approach, a novel LLM-based framework for C++ unit test generation. \approach enhances LLMs by providing a comprehensive understanding of the dependency relationships within the project under test via program analysis. Furthermore, \approach incorporates language-specific knowledge about C++ derived from project documentation and empirical observations, significantly improving the correctness of the LLM-generated unit tests. We implement \approach by employing the widely popular LLM GPT-4o. The experimental results show that \approach outperforms current state-of-the-art approaches on a collection of ten popular C++ projects. Our findings demonstrate the effectiveness of \approach in generating high-quality C++ unit tests.
\end{abstract}



\received{20 February 2007}
\received[revised]{12 March 2009}
\received[accepted]{5 June 2009}

\maketitle

\section{Introduction}
\label{int}

The C++ programming language, widely renowned for its efficiency, scalability, and security, plays a crucial role in developing basic software such as operating systems and compilers. As these systems increase in size, however, ensuring code reliability becomes increasingly challenging in the fast-evolving software development process \cite{othmane2014extending}. Unit testing \cite{runeson2006survey,daka2014survey} serves as a fundamental technique in this pursuit, providing a robust means of validating individual software units in isolation from the rest of the system. By independently verifying each unit, developers can detect and rectify defects early in the software development lifecycle, thereby improving overall code quality. Nevertheless, writing high-quality C++ unit test cases becomes a challenging and time-consuming task for developers in practice when the complexity of software systems grows \cite{garousi2013survey}. More recently, the application of large language models (LLMs) in unit test case generation has been extensively explored in both academia and industry \cite{schafer2024empirical,yuan2024evaluating,tang2024chatgpt,alshahwan2024observation,alshahwan2024automated}. For instance, Yuan et al. \cite{yuan2024evaluating} conducted a comprehensive evaluation of using ChatGPT in automatically generating unit test cases for Java projects via both quantitative analysis and user studies. Their findings indicate that the ChatGPT-generated test cases exhibit commendable readability, thereby affirming the feasibility of an LLM-based technological approach.

\begin{figure}[tbp]
  \centering
  \includegraphics[width=\textwidth]{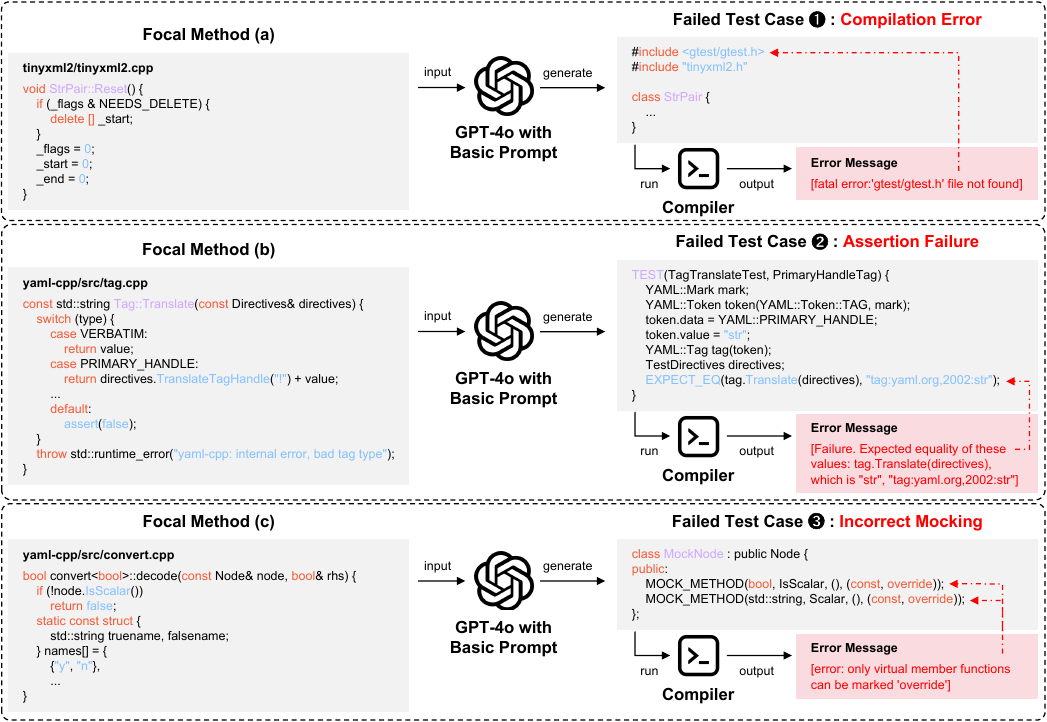}
  \caption{Limitations of LLMs in Generating C++ Unit Test Cases.}
  \Description{}
  \label{limitations}
\end{figure}

Although LLM-based approaches have achieved remarkable performance in unit test case generation, the test code generated by LLMs still encounters issues, including compilation errors and assertion failures. Furthermore, current state-of-the-art approaches and tools \cite{sapozhnikov2024testspark,ryan2024code,chen2024chatunitest,yuan2024evaluating} primarily concentrate on interpreted programming languages such as Java and Python, with limited mature solutions yet available for generating unit test cases tailored to compiled programming language like C++. Particularly, the intricate language features of C++ present substantial challenges in utilizing LLMs to generate executable unit test cases with high coverage. To underscore the limitations of LLMs in the generation of C++ unit test cases, Figure~\ref{limitations} illustrates three motivating examples that demonstrate typical scenarios in which the most advanced LLM GPT-4o, when provided with the basic prompt\footnote[1]{Following ChatTester \cite{yuan2024evaluating}, the basic prompt comprises the task description for unit test generation, a requirement to understand program's intent, the source code of the focal method, and the dependency contexts within the focal class file.}, fails to generate correct unit test cases for the corresponding focal methods {(i.e., the methods under test)} in the open-source C++ projects.

\begin{itemize}
    \item \textbf{Limitation 1: Missing code-agnostic contexts pertaining to the project configuration dependencies for LLMs to generate compilable code.} Existing LLM-based approaches \cite{chen2024chatunitest,yuan2024evaluating} leverage static analysis techniques to extract code dependency contexts related to the focal method from the focal class file, which aims to ensure the syntactic correctness of LLM-generated test code. However, LLMs may also produce compilation errors due to a lack of awareness regarding dependencies specified within the project's configuration. As shown in the \textbf{Error Message} output by \textbf{Failed Test Case \ding{202}}, the test code generated by GPT-4o fails to compile because the configuration file for the \texttt{tinyxml2} project does not include the usage of the third-party library \texttt{gtest}. Given the absence of such code-agnostic contexts (e.g., the availability of specific programming framework) in the provided basic prompt, there is a high likelihood that GPT-4o will default to directly using the \texttt{GoogleTest} framework to generate C++ unit test cases for \texttt{tinyxml2}.
    \item \textbf{Limitation 2: Lack of cross-file intended behavior in the corresponding project as guidance for LLMs to generate correct assertions.} Considering the example of \textbf{Focal Method (b)}, the functionality of the \texttt{Translate} method in the \texttt{tag} class file is to invoke the relevant interface methods provided by the \texttt{Directives} class to translate various YAML tags in accordance with their respective types. In this instance, the variable \texttt{type} is defined within the header file of \texttt{tag}, while the contextual information related to the invoked interface methods resides in the class file of \texttt{Directives}. Given that these cross-file dependency contexts within the \texttt{yaml-cpp} project are not encompassed in the basic prompt, comprehending the functional intent of \texttt{Translate} alone does not sufficiently assist GPT-4o in grasping the usage of \texttt{Translate}. Consequently, \textbf{Failed Test Case \ding{203}} is unable to configure the correct YAML tag type to align with the target translation pattern. This misalignment causes the actual outcome to deviate from the expected result, consequently triggering an assertion failure.
    \item \textbf{Limitation 3: LLMs struggle to correctly generate complex test code without an understanding of C++ language-specific domain knowledge.} In practice, the focal method may need to interact with complex data objects defined in dependent class files. In such cases, developers typically create virtual objects using mocking techniques \cite{saff2004mock,thomas2002mock} to simulate the behavior of real external dependencies. However, to effectively generate complex test code, LLMs must grasp the principles and applicable scenarios related to mocking \cite{spadini2019mock,spadini2017mock}. In \textbf{Focal Method (c)}, \texttt{decode} needs to return a boolean value by verifying whether \texttt{node} is a scalar. As illustrated in \textbf{Failed Test Case \ding{204}}, GPT-4o attempts to simulate a \texttt{Node} class object to mimic the test input. The \textbf{Error Message} indicates that the methods invoked on \texttt{node} are non-virtual functions. As a consequence, GPT-4o generates test code that contains compilation errors, stemming from an insufficient understanding of such complex language feature of C++ mocking.
\end{itemize}

To tackle the aforementioned limitations, we propose \approach, a novel framework designed to enhance the capabilities of LLMs in generating high-quality \textbf{\underline{C}}++ un\textbf{\underline{I}}t \textbf{\underline{T}}est cases by providing project-dependenc\textbf{\underline{Y}} a\textbf{\underline{WA}}reness and \textbf{\underline{L}}anguage-specific \textbf{\underline{K}}nowledge. Our key insight lies in \textit{enabling LLMs to act as skilled developers, leveraging specialized knowledge and global comprehension of project dependencies to perform unit testing effectively}. The main ideas of \approach are outlined as follows:
\begin{itemize}
    \item \textbf{Empowering LLMs with an awareness of project-level dependency relationships through program analysis.} The project dependencies employed by \approach are categorized into two types: cross-file data dependencies and configuration dependencies. To conduct cross-file data dependency analysis, \approach initially performs static analysis to identify the header and source code files within the project that exhibit a dependency-chain relationship with the focal class file. Subsequently, \approach utilizes the abstract syntax tree (AST) to extract the relevant cross-file data dependency contexts associated with the focal method from these identified files. Furthermore, our observations indicate that when the project under test necessitates specific versions of the compilation environment or third-party libraries, the probability of encountering compilation errors in the LLM-generated unit tests increases substantially. By analyzing project's configuration files, \approach gathers critical information regarding the third-party library usage and the requirements of the compilation environment. These configuration dependencies are explicitly provided to LLMs to reduce the occurrence of compilation errors.
    \item \textbf{Augmenting LLMs with intention-guiding information related to the focal method via a hybrid retrieval strategy.} Software repositories often contain extensive documentation, such as requirements specifications and API references \cite{maalej2013patterns}. Utilizing this natural language information can further aid LLMs in understanding the functional logic of the focal method. Moreover, when the focal method has long dependency chains or complex initialization of dependent objects, incorporating relevant code snippets, especially those that invoke the focal method or demonstrate its initialization, into the prompts can greatly help guide LLMs in understanding the real-world usage patterns of the focal method, thereby compensating for the limitations of static analysis techniques. To facilitate the retrieval of both natural language documentation and programming language code snippets, \approach employs a hybrid strategy designed to efficiently extract intention-guiding information from bimodal sources, ensuring an accurate and contextually relevant retrieval process.
    \item \textbf{Prompting LLMs with step-by-step instructions and language-specific knowledge derived from empirical observations.} \approach decomposes the unit test generation task into smaller, more specific subtasks with structured step-by-step instructions. The goal is to streamline the generation of C++ unit test cases by providing LLMs with detailed procedural guidance. Within \approach, LLMs are utilized to: (1) infer both the intention and dependencies of the focal method; (2) generate an entire test file for the focal method by leveraging the provided guidance; and (3) refine the generated test cases using language-specific knowledge. To perform Step (3), we further conduct an empirical analysis of the LLM-generated failed test cases for building language-specific domain knowledge tailored to C++. This knowledge-driven approach assists in guiding LLMs to emulate experienced developers in generating accurate test cases.
\end{itemize}

We implement a prototype of \approach using GPT-4o and evaluate it on a collected benchmark comprising 1288 focal methods from ten C++ projects. We conduct a comparative analysis of \approach against seven state-of-the-art baselines. Our evaluation demonstrates that \approach outperforms all baselines in both compilation success rate and code coverage. Our contributions can be summarized as follows:
\begin{itemize}
  \item We present the first attempt at enhancing the capabilities of LLMs for C++ unit test generation by leveraging project-dependency awareness and language-specific knowledge, enabling LLMs to function like human developers in writing correct unit test cases.
  \item We thoroughly evaluate \approach on a diverse collection of open-source projects, thereby substantiating the effectiveness of each component within \approach. Furthermore, we illustrate its capacity to generalize in generating high-quality unit tests across various LLMs.
  \item  We publicly release the artifacts \cite{anonymous} of \approach on Zenodo to facilitate the reproduction.
\end{itemize}

\textit{Article Organization.} The remainder of this paper is organized as follows: Section~\ref{met} introduces in detail the proposed framework. Section~\ref{exp} and Section~\ref{res} provide the experimental setups and results of our research. Section~\ref{dis} presents multi-perspective discussions and discloses the threats to the validity of our approach. Section~\ref{rel} describes the related work. Section~\ref{con} concludes with research findings and future directions.

\section{Methodology}
\label{met}

\begin{figure}[htbp]
  \centering
  \includegraphics[width=0.97\textwidth]{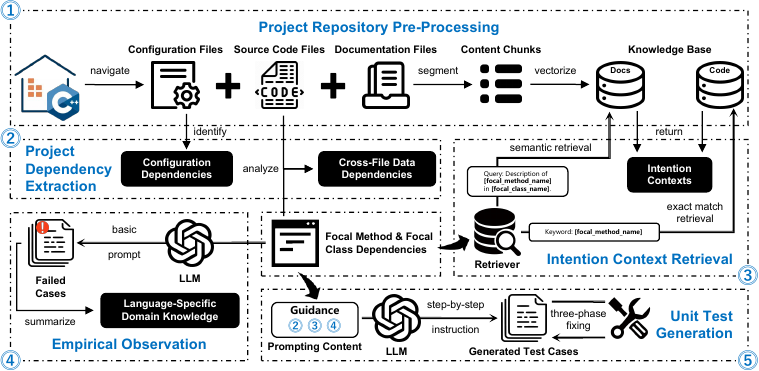}
  \caption{Overview of \approach.}
  \Description{}
  \label{overview}
\end{figure}

In this paper, we introduce \approach, a novel LLM-based framework for C++ unit test generation that addresses the limitations outlined in Section~\ref{int}. As illustrated in Figure~\ref{overview}, \approach consists of five stages designed to enhance LLMs in generating high-quality C++ unit test cases. It effectively integrates the strengths of program analysis with retrieval-augmented strategies, empowering LLMs with developer-like capabilities through three key aspects of guidance.

\subsection{Task Formulation}

We first provide a formal definition of the unit test generation task. Let $\mathtt{Repo=\{SC,C,D\}}$ be the project repository, which includes a set of source code files $\mathtt{SC=\{sc_1,sc_2,...sc_i\}}$, configuration files $\mathtt{C=\{c_1,c_2,...c_j\}}$, and documentation files $\mathtt{D=\{d_1,d_2,...d_k\}}$. The methods within $\mathtt{Repo}$ are denoted as $\mathtt{M=\bigcup _{sc \in SC}M(sc)}$, where $\mathtt{M(sc)=\{m_1,m_2,...,m_n\}}$ is the set of methods in $\mathtt{sc}$. Given a focal method $\mathtt{m_{focal} \in M(sc_{focal})}$ associated with the focal class file $\mathtt{sc_{focal}}$, the objective of \approach is to generate unit test case(s) $\mathtt{TC}$ for $\mathtt{m_{focal}}$ through the following stages:
\begin{description}
    \item[{\color{light-blue}\textbf{\ding{192} Project Repository Pre-Processing:}}] This stage involves navigating through all files in $\mathtt{Repo}$, segmenting the selected files into chunks, and storing the vectorized chunks in the knowledge bases $\mathtt{KB_{code}}$ and $\mathtt{KB_{docs}}$.
    \item[{\color{light-blue}\textbf{\ding{193} Project Dependency Extraction:}}] In this stage, we identify configuration dependencies within $\mathtt{C}$ and cross-file data dependencies between $\mathtt{sc_{focal}}$ and $\mathtt{SC}$ via program analysis.
    \item[{\color{light-blue}\textbf{\ding{194} Intention Context Retrieval:}}] This stage focuses on retrieving relevant code snippets from $\mathtt{KB_{code}}$ and documentation from $\mathtt{KB_{docs}}$ as intention contexts utilizing a hybrid strategy.
    \item[{\color{light-blue}\textbf{\ding{195} Empirical Observation:}}] In this stage, we analyze the LLM-generated failed test cases and summarize the language-specific knowledge derived from empirical observations.
    \item[{\color{light-blue}\textbf{\ding{196} Unit Test Generation:}}] This stage involves prompting the LLM with step-by-step instructions to generate test cases $\mathtt{TC \leftarrow LLM(PROMPT)}$, where $\mathtt{PROMPT}$ includes the basic prompt supplemented with additional guidance from \textbf{Stage} {\color{light-blue}\textbf{\ding{193}}}-{\color{light-blue}\textbf{\ding{195}}}, and incorporates a three-phase post-processing approach for error-fixing.
\end{description}

\subsection{Project Repository Pre-Processing}

In the context of program analysis-based and retrieval-augmented unit test generation, this stage involves two preliminary processes for pre-processing the given focal method and corresponding project. These processes are essential to help LLMs understand the code structure and its dependencies, thereby facilitating precise analysis and effective retrieval to support test case generation.

\subsubsection{Structured Focal Context Extraction}

\begin{figure}[htbp]
  \centering
  \includegraphics[width=\textwidth]{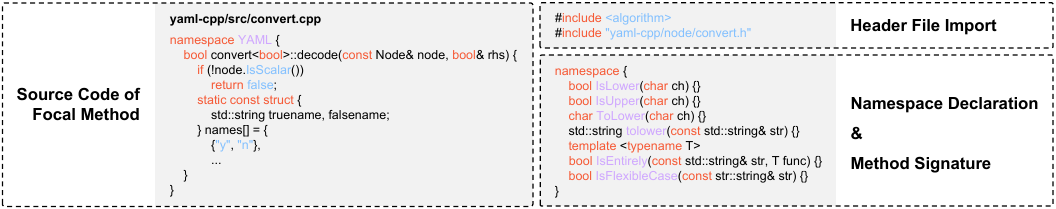}
  \caption{Illustrative Example of Structured Focal Context for the \texttt{decode} Method in the \texttt{convert} Class.}
  \Description{}
  \label{example1}
\end{figure}

Given the input focal class file, \approach begins by parsing the source code into an AST using Clang\footnote[2]{https://clang.llvm.org}. For each focal method, \approach then gathers the essential focal contexts in a structured format, providing the foundational information necessary for project dependency analysis. Figure~\ref{example1} illustrates the structured focal contexts for \texttt{decode} in the \texttt{convert} class, preserving the complete code implementation of the focal method, along with imports from C++ standard libraries, third-party libraries, user-defined header files, namespace declarations, and signatures of other methods within the focal class.

\subsubsection{Knowledge Base Construction}
\label{kbc}

Retrieval-augmented generation (RAG) allows LLMs to leverage information from external knowledge bases for reducing hallucinations \cite{chen2024benchmarking}. To enhance the LLM's understanding of the focal method during unit test generation, \approach constructs knowledge bases using the documentation and source code from the project repository. {First, \approach scans all semantically relevant files within the repository, including project documentation (e.g., requirement specifications) written in natural language and source code files in programming languages. Files that do not contribute semantic knowledge (e.g., configuration files) are excluded from constructing knowledge bases.} \approach then applies text segmentation strategies to slice the documentation and code files. For instance, markdown files can be segmented into text chunks based on title levels, while source code files are divided into method-level code chunks. Additionally, natural language comments within the code are extracted separately as text chunks. To facilitate efficient semantic retrieval and similarity calculations, \approach utilizes an embedding model BGE \cite{xiao2024cpack} for the vectorization of segmented text chunks, storing the resulting vectors in $\mathtt{KB_{docs}}$. The segmented code chunks are directly stored in $\mathtt{KB_{code}}$ for exact match retrieval without vectorization.

\subsection{Project Dependency Extraction}
\label{pde}

\begin{figure}[tbp]
  \centering
  \includegraphics[width=\textwidth]{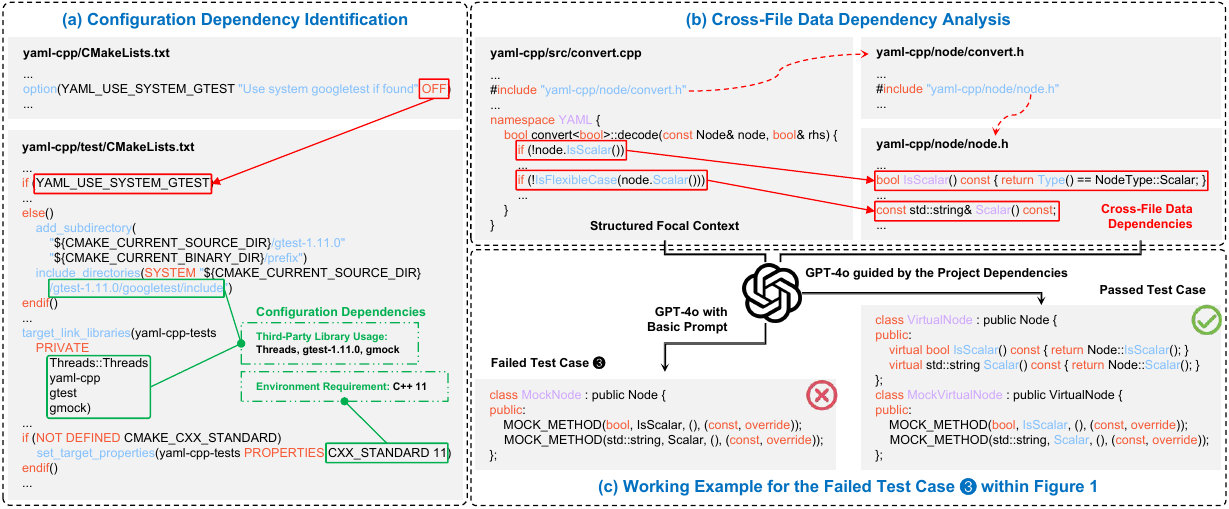}
  \caption{Illustrative Example of Project Dependency Extraction.}
  \Description{}
  \label{example2}
\end{figure}

LLM-generated test code often fails to compile, typically due to missing dependencies or specific version requirements of the compilation environment. This limitation compels LLMs to make assumptions about the usage of undefined variables or methods based on their reasoning \cite{zhang2024llm}. To mitigate compilation errors, \approach enables LLMs to be aware of relevant dependency contexts at the project level through program analysis. The project dependencies extracted by \approach are classified into the following two categories:
\begin{itemize}
    \item \textbf{Configuration Dependencies:} As illustrated in Figure~\ref{example2}(a), \approach identifies both the {\color{light-green}\textbf{usage of third-party libraries}} and the {\color{light-green}\textbf{compilation environment requirements}} of the project under test. \approach first parses the \texttt{CMakeLists} files in the \texttt{root} and \texttt{test} dictionaries using regular expressions to extract configuration dependencies based on keyword-matching, including the command like {\color{light-blue-1}\texttt{target\_link\_libraries}}. This process captures details about dependent third-party libraries and their versions (e.g., \textbf{gtest-1.11.0}) as well as the compilation environment requirement (i.e., \textbf{C++ 11}) specified through {\color{light-blue-1}\texttt{set\_target\_properties}}. The parsed dependencies (framed by the {\color{light-green}green rectangle}) are stored in a structured format that facilitates seamless integration with LLM prompts. This enables LLMs to understand the necessary code-agnostic dependencies of the project under test, which is essential for generating compilable test code.
    \item \textbf{Cross-File Data Dependencies:} In addition to the focal contexts utilized by existing LLM-based approaches \cite{chen2024chatunitest,yuan2024evaluating}, \approach conducts data dependency analysis via AST to gather cross-file dependencies as contextual guidance, helping LLMs in reasoning accurately with sufficient information. \approach first navigates all files with specific suffixes (i.e., ".cpp" and ".h"). In C++, user-defined header files are typically imported using \texttt{\#include ""}. Thus, \approach utilizes program analysis to filter out files with direct dependencies on the focal class file based on the \texttt{include} field information. This analysis process is recursive, and \approach limits the extracted dependency chain to two layers. {This design is primarily motivated by the context window constraints of employed LLMs, ensuring that the length of extracted contextual dependencies remains within the model's processing capacity.} Subsequently, \approach employs data-flow analysis on the AST nodes of the invoked methods within the focal method to extract cross-file data dependency contexts from the filtered files. As shown in Figure~\ref{example2}(b), \approach extracts the dependent contexts (framed by the {\color{red}red rectangle}) for the invoked methods (i.e., \texttt{IsScalar} and \texttt{Scalar}) in \texttt{decode} from the \texttt{Node} class through a two-layer relationship (i.e., \texttt{convert.cpp} {\color{red}$\dashrightarrow$} \texttt{convert.h} {\color{red}$\dashrightarrow$} \texttt{node.h}). Furthermore, Figure~\ref{example2}(c) presents a working example of how cross-file data dependencies can guide LLMs in generating executable test code for \textbf{Failed Test Case \ding{204}} shown in Figure~\ref{limitations}. By incorporating additional contextual information from \texttt{Node} and prompting LLMs with the language-specific knowledge of \texttt{gmock}, GPT-4o derives a class and declares virtual functions, successfully simulating the behavior of the invoked methods within \texttt{decode} and generating accurate test inputs.
\end{itemize}

\subsection{Intention Context Retrieval}

To better guide LLMs in understanding the program intention of the focal method, \approach employs RAG to generate effective unit test cases by leveraging the project documentation and source code stored in the knowledge bases described in Section~\ref{kbc}. The natural language documentation provides functionality requirements of the focal method as articulated by developers, while the source code snippets within the project offer LLMs real-world examples of focal method invocation and initialization. This approach mitigates issues related to focal methods that have dependency chains longer than two layers or are complex to initialize. Specifically, \approach employs a hybrid retrieval strategy that conducts knowledge queries in two stages. For $\mathtt{KB_{docs}}$, the process begins by generating a corresponding query statement (as shown in Figure~\ref{overview}) based on the name of the focal method and the class to which it belongs. The embedding model BGE is then utilized to convert this query into a vector representation, capturing its semantic information and enabling alignment with the content of $\mathtt{KB_{docs}}$ in vector space. Using the retrieval algorithms provided by the vector database Faiss \cite{douze2024faiss}, the similarity between the query vector and the vectors in $\mathtt{KB_{docs}}$ is computed, employing cosine similarity to identify the vectors that are most similar to the query vector. Based on the similarity scores, the top-2 most relevant responses are selected as retrieval results. This setup ensures retrieval accuracy while providing sufficiently detailed guidance information, assisting in the generation of more comprehensive test cases. Preliminary practices indicate that semantic retrieval for code snippet examples often yields imprecise results \cite{bajracharya2010leveraging,yan2020code}. {To optimize retrieval from $\mathtt{KB_{code}}$, \approach adopts an exact match strategy based on the focal method's signature, including its name and parameter types, extracted through static analysis. Specifically, \approach first utilizes the regular expression: \texttt{r`\textbackslash b'} + \texttt{re.escape(focal\_method\_name)} + \texttt{r`\textbackslash s*\textbackslash([\textasciicircum)]*\textbackslash)'} to identify candidate method invocation or initialization examples from $\mathtt{KB_{code}}$. Following this, a verification step filters out candidate examples with mismatched parameter counts or types, ensuring only those that exactly match the focal method's signature are retained. For example, \texttt{focal\_method\_name(double)} will not match \texttt{focal\_method\_name(int)}. This two-step strategy guarantees accurate retrieval while effectively distinguishing between similar method variants.}

\subsection{Empirical Observation}

In addition to the extracted project dependencies and retrieved intention contexts, \approach further provides error patterns, along with their corresponding solution guidelines, as language-specific domain knowledge. This knowledge tailored to C++ programming language is used to prompt LLMs with instructions aimed at mitigating common errors in the task of unit test generation. To achieve this, we conduct an empirical study to evaluate the correctness of C++ test cases generated by existing state-of-the-art LLMs through quantitative analysis. The procedure for our empirical investigation is outlined as follows:
\begin{itemize}
    \item \textbf{Empirical Setups:} (1) \textbf{Benchmark.} We adopt a diverse set of real-world open-source C++ projects collected from GitHub, with the selection criteria and detailed statistics provided in Section~\ref{ben}. (2) \textbf{Subject LLMs.} We select two state-of-the-art LLMs to evaluate their capabilities in generating high-quality C++ unit test cases based on our collected benchmark: the open-source LLM DeepSeek-V3 \cite{deepseekai2024deepseekv3} and the closed-source LLM GPT-4o \cite{openai2023gpt4}. (3) \textbf{Prompt Design.} As suggested by Yuan et al. \cite{yuan2024evaluating}, we design our prompt by closely following common practices in recent unit test generation research. The prompt consists of a natural language description that explains the task to the LLM, along with the code context, which includes the focal method and other relevant contextual information (e.g., the fields and method signatures within the focal class).
    \item \textbf{Experimental Procedure:} {For each project in the benchmark, we clone its repository from GitHub and extract relevant information to support prompt construction.} We query the selected LLMs using our designed prompt for each focal method and consider the test cases generated by the LLMs as the output. We use the official API of each LLM with the configuration set to generate the top-1 chat completion choice and a sampling temperature of 0. The generated test cases are then placed in the test directory of the project, where we attempt to compile and execute them for subsequent analysis. In our experimental design, we evaluate the correctness of the generated test cases from three perspectives. First, we use the Clang parser as a syntax checker to verify the syntactic correctness of the generated test cases. {Next, we measure the correctness of compilation and execution by verifying whether the generated test cases compile successfully and run without errors. Error messages produced during compilation and execution are automatically extracted for analysis. To examine the failed test cases generated by the two evaluated LLMs, the first two authors manually classified the errors based on the corresponding compiler-generated messages. This empirical process involved over 1000 failed test cases and required approximately five hours. To ensure consistency and accuracy, the two authors collaboratively reviewed and resolved discrepancies through discussion, reaching consensus on all classification results.} In addition, we employ \texttt{llvm-cov}\footnote[3]{https://llvm.org/docs/CommandGuide/llvm-cov.html} to collect both line and branch coverage, providing an assessment of the sufficiency of the generated test cases.
    \item \textbf{Error Analysis:} To better understand the limitations of existing LLMs in C++ unit test generation, we analyze common error patterns in the failed test cases generated by DeepSeek-V3 and GPT-4o{, with a particular focus on their impact on compilation correctness. It is important to note that this analysis is limited to \textbf{compilation errors}, which constitute a significant proportion of all errors and represent a critical initial barrier to the successful execution and correctness of generated test cases. Specifically, we automatically categorize each failed test case based on the associated compilation error message. Table~\ref{ceb} presents the distribution of compilation errors observed in the test cases generated by the evaluated LLMs. The ``Frequency'' column reports the number of occurrences for each error pattern across \textbf{DeepSeek-V3}, \textbf{GPT-4o}, and their combined total. For clarity, only error patterns that appeared more than ten times are included in the table. As shown in Table~\ref{ceb}, LLM-generated test cases exhibit a diverse range of compilation error patterns. The most common errors are related to undefined symbols, typically caused by references to unresolved identifiers such as undeclared methods or variables. Two other prevalent error patterns include access violations and type mismatches. Access errors generally result from invalid attempts to reference class members, while type errors stem from incompatible expressions or incorrect type assignments. Additionally, we observe that \textbf{GPT-4o} frequently produces test code with incorrect namespace usage, which is not present in \textbf{DeepSeek-V3} outputs. This quantitative analysis reveals persistent patterns of compilation errors encountered by LLMs during unit test generation and provides valuable insights that inform the design of our mitigation strategies.}
    
    \begin{table}[t]
    \centering
    \caption{{Compilation Error Breakdown for LLM-Generated Test Cases}}
    \label{ceb}
    \begin{tabular}{llccc}
        \toprule
        \multirow{2.5}{*}{{\textbf{Error Pattern}}} & \multirow{2.5}{*}{{\textbf{Error Description}}} & \multicolumn{3}{c}{{\textbf{Frequency}}} \\
        \cmidrule[0.5pt](rl){3-5}
        & & {\textbf{DeepSeek-V3}} & {\textbf{GPT-4o}} & {\textbf{Total}} \\
        \midrule
        {Undefined Symbols Error} & {Missing or unresolved identifiers} & {351} & {382} & {733} \\
        {Access Error} & {Invalid access to class members} & {157} & {147} & {304} \\
        {Type Error} & {Type mismatches in expressions or assignments} & {140} & {161} & {301} \\
        {Other} & {Miscellaneous unclassified errors} & {91} & {76} & {167} \\
        {Test Setup Error} & {Failure during initialization of tests} & {51} & {99} & {150} \\
        {Linker Error} & {Cross-file linkage failure} & {92} & {32} & {124} \\
        {Syntax Error} & {Invalid syntax in C++ source files} & {18} & {59} & {77} \\
        {Namespace Error} & {Incorrect or missing namespace usage} & {0} & {45} & {45} \\
        {Multiple Definition Error} & {Duplicate symbols defined in different files} & {5} & {26} & {31} \\
        {Template Error} & {Invalid usage of C++ templates} & {1} & {9} & {10} \\
        \bottomrule
    \end{tabular}
\end{table}

\begin{table}[t]
    \centering
    \caption{Language-Specific Domain Knowledge Derived from Empirical Observations}
    \label{knowledge}
    \resizebox{\textwidth}{!}{
    \begin{tabular}{ll}
        \toprule
        \multicolumn{1}{c}{\textbf{Category}} & \multicolumn{1}{c}{\textbf{Guideline}} \\
        \midrule
        \multirow{6}{*}{\textbf{(A) Compilation Error}} & (A.1) Import all necessary dependencies with the correct paths. \\
        & (A.2) Use only the C++ standard libraries, imported third-party libraries, and provided methods. \\
        & (A.3) If \texttt{gtest} is not allowed, directly call test methods from the main function. \\
        & (A.4) Use the correct namespace throughout the tests. \\
        & (A.5) Properly handle static members by accessing them using the class name. \\
        & (A.6) Avoid invoking private methods or accessing private fields defined in the program. \\
        \midrule
        \multirow{2}{*}{\textbf{(B) Execution Failure}} & (B.1) Choose appropriate assertions for the pointer data type, clearly distinguishing between address and content comparisons. \\
        & (B.2) For mocking (if using \texttt{gmock}), remember that only virtual methods can be mocked. \\
        \midrule
        \multirow{2}{*}{\textbf{(C) Poor Coverage}} & (C.1) Ensure coverage of true and false branches for each conditional predicate at least once. \\
        & (C.2) Utilize non-terminating assertions (e.g., \texttt{EXPECT\_*}) to maximize code coverage. \\
        \bottomrule
    \end{tabular}}
\end{table}
    
    \item \textbf{Solution Guidelines:} As presented in the first column of Table~\ref{knowledge}, we manually group similar errors into three high-level categories: \textbf{(A) Compilation Errors}, \textbf{(B) Execution Failures}, and \textbf{(C) Poor Coverage}. While compilation errors constitute the majority, the LLM-generated test cases also exhibit execution failures, often caused by incorrect assertions or flawed mocking code. Furthermore, poor coverage emerges as a significant issue, largely attributable to the LLMs' limited ability to exercise all conditional branches and their misuse of assertion types. As listed in the second column of Table~\ref{knowledge}, we systematically summarize solution guidelines for the corresponding error patterns based on empirical observations of the failed test cases generated by the selected LLMs. As discussed in Section~\ref{pde}, \textbf{Failed Test Case \ding{204}} can be addressed by leveraging additional cross-file data dependencies along with \textbf{Guideline (B.2)}. Specifically, we organize \textbf{Guideline (B.2)} as instructions to prompt LLMs in a knowledge-driven manner for tackling the incorrect mocking issue. In doing so, \approach incorporates language-specific knowledge about mocking in C++, including best practices for creating mock objects for virtual functions. In the case of the \texttt{decode} method, \approach provides LLMs with detailed guidance on how to correctly mock non-virtual member functions in C++. By enhancing the LLM's understanding of these language-specific nuances, \approach ensures the generated mocking code is both syntactically and semantically correct, effectively preventing execution failures caused by improper mocking of non-virtual methods. While these guidelines do not cover all error types, they are instrumental in guiding LLMs to produce high-quality test cases and mitigate common errors in C++ unit test generation.
\end{itemize}

\subsection{Unit Test Generation}

Unit test generation is a complex task that poses significant challenges when generating high-quality test cases from scratch with limited guidance. To address this, {\approach initially produces an entire test file for the given focal method by following step-by-step instructions.} Subsequently, \approach corrects the failed test cases using a three-phase rule-based fixing approach. As illustrated in Algorithm~\ref{algorithm1}, we provide detailed process overview as follows.

\begin{algorithm}[htbp]
\small
	\caption{Unit Test Case Generation for a Focal Method.}
	\label{algorithm1}
	\KwIn{The given focal method: \(\mathtt{m_{focal}}\); The focal contexts: \(\mathtt{Context_{focal}}\); The configuration dependencies: \({\mathtt{Dep_{c}}}\); The cross-file data dependencies: \(\mathtt{Dep_{d}}\); The intention contexts: \(\mathtt{Context_{intent}}\); The guidelines of language-specific domain knowledge: \(\mathtt{Guideline_{DK}}\); The step-by-step instruction prompt: \(\mathtt{PROMPT_{step}}\); The syntactic error fixing rules: \(\mathtt{Rule_{F_s}}\); The compilation error fixing rules: \(\mathtt{Rule_{F_c}}\); The error fixing prompt: \(\mathtt{PROMPT_{F}}\)}
	\KwOut{The generated unit test case(s): \(\mathtt{TC}\)}
	\BlankLine
	\textit{/* Initial Unit Test Case Generation via Step-by-Step Instructions */}
	\BlankLine
	\textit{/* Step 1: Program Understanding */}
	
	\(\mathtt{Intent}\), \(\mathtt{Dep_{ingredient}}\) \(\leftarrow\) \(\mathbf{LLM}\)(\(\mathtt{PROMPT_{step}(m_{focal})}\))
	\BlankLine
	\textit{/* Step 2: Unit Test Generation */}
	
	\(\mathtt{TC}\) \(\leftarrow\) \(\mathbf{LLM}\)(\(\mathtt{PROMPT_{step}(m_{focal}, Context_{focal}, Dep_c, Dep_d, Context_{intent}, Intent, Dep_{ingredient})}\))
	\BlankLine
	\textit{/* Step 3: Test Case Refinement */}
	
	\(\mathtt{TC}\) \(\leftarrow\) \(\mathbf{LLM}\)(\(\mathtt{PROMPT_{step}(TC, Guideline_{DK})}\))
	\BlankLine
	\textit{/* Post-Processing of the LLM-Generated Unit Test Cases */}
	
	\(\mathtt{TC}\) \(\leftarrow\) \(\mathtt{Rule_{F_s}(TC)}\)
	
	\uIf{\(\mathbf{Compiler}(\mathtt{TC})\) is not \(\mathbf{PASS}\)}{\(\mathtt{TC} \leftarrow \mathtt{Rule_{F_c}(TC)}\)\\
	\uIf{\(\mathbf{Compiler}(\mathtt{TC})\) is not \(\mathbf{PASS}\)}{\(\mathtt{Context_{error} \leftarrow \mathbf{Compiler}(\mathtt{TC})}\)\\\(\mathtt{TC} \leftarrow \mathbf{LLM}\)(\(\mathtt{PROMPT_{F}(TC, Context_{error})}\)}}
	\algorithmicreturn{ \(\mathtt{TC}\) }
\end{algorithm}

\begin{figure}[tbp]
  \centering
  \includegraphics[width=\textwidth]{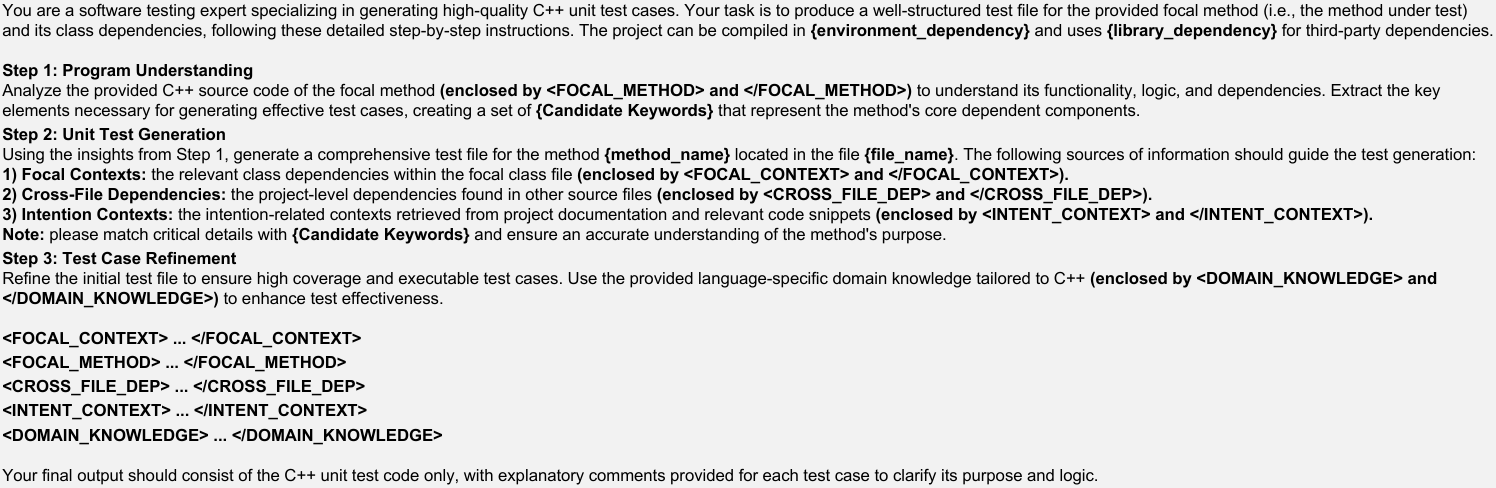}
  \caption{The Detailed Prompting Content for Generating Initial Test Cases for a Given Focal Method.}
  \Description{}
  \label{prompt}
\end{figure}

\subsubsection{Initial Unit Test Case Generation via Step-by-Step Instructions}

As shown in Figure~\ref{prompt}, the prompting content for querying the LLM comprises four components: the \textbf{task definition}, the \textbf{step-by-step instructions}, the \textbf{contextual information}, and the \textbf{output format}. Within the context of \approach, the LLM is employed to generate initial test cases by following steps: 
\begin{enumerate}
    \item \textit{Program Understanding} \textbf{(Line 3)}: First, the LLM analyzes the provided source code of \(\mathtt{m_{focal}}\) to grasp its intended functionality, denoted as \(\mathtt{Intent}\). Next, the LLM extracts the key elements necessary for generating effective test cases, creating a set of candidate keywords (i.e., \(\mathtt{Dep_{ingredient}}\)) that represent the core dependent ingredients of \(\mathtt{m_{focal}}\).
    \item \textit{Unit Test Generation} \textbf{(Line 5)}: Based on the \(\mathtt{Intent}\) of \(\mathtt{m_{focal}}\), the LLM generates an initial test file for \(\mathtt{m_{focal}}\) by utilizing the provided contextual information, which includes the focal contexts \(\mathtt{Context_{focal}}\), project dependencies \(\mathtt{Dep_{c}}\) and \(\mathtt{Dep_{d}}\), intention contexts \(\mathtt{Context_{intent}}\). Additionally, the LLM matches the contextual information with the extracted \(\mathtt{Dep_{ingredient}}\) to ensure an precise understanding of the purpose of \(\mathtt{m_{focal}}\).
    \item \textit{Test Case Refinement} \textbf{(Line 7)}: To enhance the effectiveness of the LLM-generated test cases \(\mathtt{TC}\), the LLM refines the initial test file using the language-specific domain knowledge as guidelines \(\mathtt{Guideline_{DK}}\), ensuring the generation of high coverage and executable \(\mathtt{TC}\).
\end{enumerate}

\begin{figure}[tbp]
  \centering
  \includegraphics[width=\textwidth]{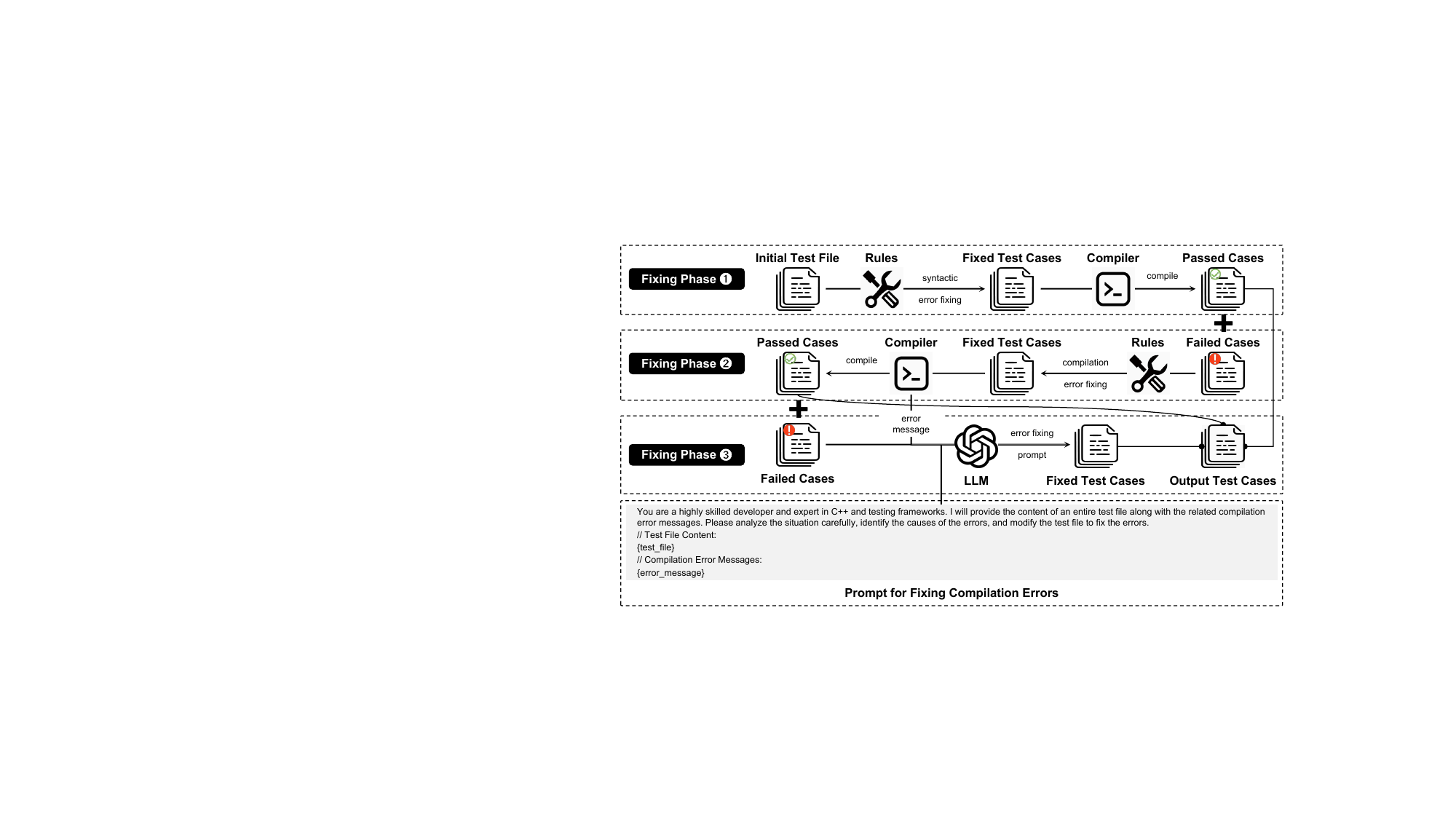}
  \caption{The Three-Phase Fixing Process for the LLM-Generated Unit Test Cases.}
  \Description{}
  \label{post}
\end{figure}

\subsubsection{Post-Processing of the LLM-Generated Unit Test Cases}
\label{postprocessing}

Inspired by ChatUniTest \cite{chen2024chatunitest}, \approach applies post-processing techniques to fix \(\mathtt{TC}\) that contain syntactic and compilation errors. As shown in Figure~\ref{post}, the fixing process consists of the following three phases: 
\begin{enumerate}
    \item[\textbf{\ding{202}}] \textbf{Rule-based Syntactic Error Fixing}: \approach first utilizes \(\mathtt{Rule_{F_s}}\) to address the syntactic errors in \(\mathtt{TC}\) \textbf{(Line 9)}. The syntactic error fixing rules include:
    \begin{itemize}
        \item Ensuring all brackets are closed through pattern matching.
        \item Removing all incorrect import statements by consolidating imports within \(\mathtt{Context_{focal}}\).
        \item If the project uses \texttt{gtest}, retaining only one \texttt{main} function in the generated test file; otherwise, ensuring each test method is called by the \texttt{main} function.
    \end{itemize}
    \item[\textbf{\ding{203}}] \textbf{Rule-based Compilation Error Fixing}: Following \textbf{Phase \ding{202}}, \approach compiles the generated \(\mathtt{TC}\) and applies \(\mathtt{Rule_{F_c}}\) to the failed cases \textbf{(Lines 10-11)}. The fixing rules include:
    \begin{itemize}
        \item Fixing incorrect usage of namespaces.
        \item Deleting non-existent import paths.
    \end{itemize}
    \item[\textbf{\ding{204}}] \textbf{LLM-based Compilation Error Fixing}: If compilation errors persist after \textbf{Phase \ding{203}}, \approach transitions to a one-round LLM-based fixing phase \textbf{(Lines 12-14)}. During this phase, \approach collects details about the failed test cases along with their associated compilation error messages to construct the fixing prompt. The LLM is then prompted to analyze the root cause of the errors and make corrections to the failed test cases. If any test case continues to fail compilation after this phase, it will be removed from \(\mathtt{TC}\).
\end{enumerate}

\section{Experimental Setup}
\label{exp}

\subsection{Research Questions}

To assess the effectiveness of \approach, we raise the following two research questions (RQs): 
\begin{itemize}
  \item \textbf{RQ1: How does \approach perform in C++ unit test generation when compared to state-of-the-art baselines?} This RQ aims to evaluate the superior effectiveness of \approach in comparison to open-source code LLMs, closed-source general-purpose LLMs, and LLM-based unit test generation approaches within the context of C++ unit test generation. To achieve this, we conduct a comprehensive evaluation of \approach against seven baselines using a collection of ten C++ projects. {Furthermore, we assess \approach's generalization capabilities using three additional LLM baselines, thereby enhancing evaluation diversity.}
  \item \textbf{RQ2: How does each component impact the performance of \approach?} Since \approach introduces a step-by-step prompting strategy and leverages various contextual information and post-processing techniques as guidance, this RQ seeks to analyze the contributions of each component through an ablation study.
\end{itemize}

\subsection{Benchmark}
\label{ben}

To comprehensively evaluate the quality of LLM-generated test cases, we construct a new benchmark consisting of ten real-world open-source C++ projects from GitHub. As shown in Table~\ref{benchmark}, the first four projects are widely utilized in recent studies on automated C++ unit testing \cite{herlim2022citrus,rho2023coyote,rho2024taming}. Additionally, we crawl two practical projects related to basic software: \texttt{ninja} (build system) and \texttt{leveldb} (database). To mitigate potential data leakage concerns, we include four projects—\texttt{json.cpp}, \texttt{glomap}, \texttt{papy}, and \texttt{mlx}—all created after the GPT-4o knowledge cutoff date (October 2023). This ensures that GPT-4o was not trained on these projects. The selection criteria for these projects are as follows. First, each project has received more than 50 stars on GitHub and is actively maintained, ensuring ongoing community interest and updates. Second, the projects span a range of application domains, including parsers (e.g., XML or YAML), basic software systems, and algorithm libraries. Third, the projects feature a number of complex methods with a cyclomatic complexity \cite{mcCabe1976complexity} greater than 10, which suggests the presence of nested control flows, while also covering intricate C++ language features. The size of the selected projects varies from 1.9K lines of code (LoC) to 149.4K LoC. Larger projects are excluded to effectively manage token costs within our limited budget. In total, \approach generates C++ unit test cases for 1288 focal methods across these ten projects. These selection criteria ensure that our benchmark is both high-quality and reproducible, while maintaining a balance between diversity and resource constraints.

\begin{table}[htbp]
    \centering
    \caption{Statistics of the Collected Open-Source C++ Projects}
    \label{benchmark}
    \resizebox{\textwidth}{!}{
    \begin{tabular}{llrrrrrc}
        \toprule
        \multicolumn{1}{c}{\textbf{Project}} & \multicolumn{1}{c}{\textbf{Application Type}} & \multicolumn{1}{c}{\textbf{GitHub Stars}} & \multicolumn{1}{c}{\textbf{Size (LoC)}} & \multicolumn{1}{c}{\textbf{\# Files}} & \multicolumn{1}{c}{\textbf{\# Focal Methods}} & \multicolumn{1}{c}{\textbf{\# Complex Methods}} & \textbf{Trained?} \\
        \midrule
        hjson-cpp\footnotemark[4] & User Interface for JSON & 73 & 2911 & 4 & 25 & 9 (36.0\%) & Y \\
        tinyxml2\footnotemark[5] & XML Parser & 5.4K & 3606 & 1 & 158 & 5 (3.2\%) & Y \\
        yaml-cpp\footnotemark[6] & YAML Parser and Emitter & 5.6K & 8800 & 28 & 204 & 31 (15.2\%) & Y \\
        re2\footnotemark[7] & Regular Expression Engine & 9.4K & 20373 & 10 & 146 & 41 (28.1\%) & Y \\
        ninja\footnotemark[8] & Build System & 12.2K & 37512 & 19 & 238 & 38 (16.0\%) & Y \\
        leveldb\footnotemark[9] & Key-Value Storage Library & 38K & 149371 & 17 & 220 & 6 (2.7\%) & Y \\
        \midrule
        json.cpp\footnotemark[10] & JSON Parsing Library & 747 & 62677 & 1 & 34 & 4 (11.8\%) & N \\
        glomap\footnotemark[11] & Map Management Library & 1.9K & 8477 & 9 & 32 & 4 (12.5\%) & N \\
        {papy\footnotemark[12]} & {JSON Data Generator} & {62} & {1869} & {6} & {25} & {2 (8.0\%)} & {N} \\
        {mlx\footnotemark[13]} & {Array Framework} & {21.9K} & {20137} & {11} & {206} & {18 (8.7\%)} & {N} \\
        \bottomrule
    \end{tabular}}
\end{table}

\footnotetext[4]{https://github.com/hjson/hjson-cpp}
\footnotetext[5]{https://github.com/leethomason/tinyxml2}
\footnotetext[6]{https://github.com/jbeder/yaml-cpp}
\footnotetext[7]{https://github.com/google/re2}
\footnotetext[8]{https://github.com/ninja-build/ninja}
\footnotetext[9]{https://github.com/google/leveldb}
\footnotetext[10]{https://github.com/jart/json.cpp}
\footnotetext[11]{https://github.com/colmap/glomap}
\footnotetext[12]{{https://github.com/noahpop77/Papy}}
\footnotetext[13]{{https://github.com/ml-explore/mlx}}

\subsection{Baselines}
\label{bas}

This paper focuses on addressing the C++ unit test generation task using LLMs. To this end, we compare \approach against seven state-of-the-art baselines. First, we select two representative open-source code LLMs: \textbf{CodeGeeX4} \cite{zheng2023codegeex} and \textbf{DeepSeek-V3} \cite{deepseekai2024deepseekv3}, both of which have demonstrated competitive performance on recent public benchmarks related to coding tasks, such as BigCodeBench \cite{zhuo2024bigcodebench} and NaturalCodeBench \cite{zhang2024naturalcodebench}. The selection of these two models is motivated by their distinct characteristics in terms of parameter size (CodeGeeX4 has 9B parameters, while DeepSeek-V3 comprises 671B parameters) and architecture (CodeGeeX4 utilizes a Transformer-based architecture, whereas DeepSeek-V3 employs a Mixture of Experts-based architecture). We exclude CodeLLaMA \cite{roziere2023code} from our evaluation due to its limited context length. Additionally, we include the closed-source general-purpose LLMs (i.e., \textbf{{GPT-3.5}} \cite{openai2022chatgpt} and \textbf{GPT-4o} \cite{openai2023gpt4}) because of their established effectiveness across a broad spectrum of tasks. {Finally, we compare \approach with three LLM-based unit test generation approaches for Java and JavaScript: \textbf{ChatTester} \cite{yuan2024evaluating}, \textbf{HITS} \cite{wang2024hits}, and \textbf{\testpilot} \cite{schafer2024empirical}}. Since existing automated C++ unit test generation tools, such as Coyote \cite{rho2023coyote} and CITRUS \cite{herlim2022citrus}, are currently unavailable, we are unable to reproduce the results reported in their respective papers. Therefore, we do not include them as baselines in this paper.

\subsection{Metrics}

As illustrated in Algorithm~\ref{algorithm1}, \approach generates a test file for one focal method at a time. It executes the generated files individually and utilizes \texttt{llvm-cov} to compute coverage for each focal method. Specifically, this paper employs four evaluation metrics commonly used in existing studies \cite{yuan2024evaluating,chen2024chatunitest,yang2024evaluation} to compare the performance of \approach with the baselines. These metrics collectively provide a comprehensive perspective on the quality, completeness, and effectiveness of the LLM-generated test cases.
\begin{itemize} 
    \item \textbf{Compilation Success Rate (CSR)}: This metric represents the percentage of LLM-generated test cases that compile successfully relative to the total test case number.
    \item \textbf{Execution Pass Rate (EPR)}: This metric quantifies the percentage of LLM-generated test cases that pass during execution, reflecting the proportion of tests that yield expected outcomes.
    \item \textbf{Line Coverage ($\mathbf{Cov_L}$)}: This metric assesses the percentage of source code lines within the focal methods that are executed by LLM-generated test cases.
    \item \textbf{Branch Coverage ($\mathbf{Cov_B}$)}: This metric evaluates the percentage of logical conditions in the source code that are explored by LLM-generated test cases.
\end{itemize}

\subsection{Implementation}
\label{imp}

We implement the core logic of \approach in Python, invoking GPT-4o via its API, specifically using the \textbf{gpt-4o} version from the GPT family of models, which is recognized as the most advanced model currently available. We do not truncate input prompts, as the selected LLMs can effectively handle lengthy inputs. We set a maximum output limit of 4096 tokens. In all experiments, we utilize greedy decoding to generate responses, with \approach producing the top-1 chat completion choice for each input prompt. To enhance response stability, we set the sampling temperature to 0. Additionally, we conduct experiments in a zero-shot setting, where no task examples are provided, thereby demonstrating the superiority of \approach.

\section{Results and Analysis}
\label{res}

\subsection{Answering RQ1}

To answer this question, we conduct a comprehensive comparison of \approach against seven baselines using the collected benchmark. For each LLM, we utilize its inference API for implementation, employing the same basic prompt described in Section~\ref{int}. {Specifically, we employ the \textbf{gpt-3.5-turbo-0125} version of the {GPT-3.5} model.} For ChatTester \cite{yuan2024evaluating}, HITS \cite{wang2024hits}, and \testpilot \cite{schafer2024empirical}, our implementation is based on their open-source reproduction artifacts from GitHub. Furthermore, we apply the same configuration settings as those used in \approach for fair comparison\footnote[14]{{Note that the maximum prompt length used in our experiments (14733 tokens) remains within the context window limits of all evaluated LLMs—{GPT-3.5} (16K), CodeGeeX4 (128K), DeepSeek-V3 (128K), and GPT-4o (128K). Therefore, no prompt truncation is necessary during any of the experiments.}}.

\begin{table}[tbp]
\centering
    \caption{Comparison of \approach against the Baselines in Terms of \textbf{Compilation Success Rate (CSR)}}
    \label{rq1a}
    \resizebox{\textwidth}{!}{
    \begin{tabular}{lrrrrrrrrrrr}
    \toprule
    \textbf{Project} & \multicolumn{1}{c}{hjson-cpp} & \multicolumn{1}{c}{tinyxml2} & \multicolumn{1}{c}{yaml-cpp} & \multicolumn{1}{c}{re2} & \multicolumn{1}{c}{ninja} & \multicolumn{1}{c}{leveldb} & \multicolumn{1}{c}{json.cpp} & \multicolumn{1}{c}{glomap} & \multicolumn{1}{c}{{papy}} & \multicolumn{1}{c}{{mlx}} & \multicolumn{1}{c}{\textbf{Avg.}} \\
    \midrule
    \textbf{CodeGeeX4} & 0.00\% & 6.95\% & \underline{38.98\%} & 0.09\% & 4.22\% & 11.75\% & 10.70\% & 35.75\% & {50.00\%} & {51.35\%} & 20.98\% \\
    \textbf{DeepSeek-V3} & 0.00\% & 0.00\% & 17.79\% & 0.00\% & 7.35\% & 13.55\% & 25.14\% & \underline{51.93\%} & {18.75\%} & {68.34\%} & 20.29\% \\
    \textbf{{GPT-3.5}} & 0.00\% & 0.00\% & 19.34\% & \underline{5.45\%} & 5.00\% & 9.64\% & 31.16\% & 28.57\% & {52.22\%} & {72.80\%} & 22.42\% \\
    \textbf{GPT-4o} & 16.17\% & 41.26\% & 22.01\% & 0.00\% & 6.90\% & 9.74\% & 51.62\% & 49.70\% & {47.90\%} & {\underline{73.07\%}} & \underline{31.84\%} \\
    \textbf{ChatTester} & 20.00\% & 18.02\% & 6.42\% & 4.68\% & 11.36\% & 0.00\% & 18.75\% & 8.82\% & {20.00\%} & {18.01\%} & 12.61\% \\
    \textbf{HITS} & 1.10\% & 5.94\% & 5.65\% & 0.00\% & \underline{20.37\%} & 2.66\% & 0.00\% & 1.92\% & {0.00\%} & {0.00\%} & 3.76\% \\
    {\textbf{\testpilot}} & {\underline{26.67\%}} & {\underline{69.23\%}} & {20.47\%} & {0.00\%} & {14.22\%} & {\underline{26.18\%}} & {\underline{70.68\%}} & {28.00\%} & {\underline{54.17\%}} & {2.66\%} & {31.23\%} \\
    \midrule
    \textbf{\approach} & \textbf{100.00\%} & \textbf{97.25\%} & \textbf{80.28\%} & \textbf{64.85\%} & \textbf{70.09\%} & \textbf{47.83\%} & \textbf{100.00\%} & \textbf{96.70\%} & {\textbf{84.29\%}} & {\textbf{92.61\%}} & \textbf{83.39\%} \\
    \bottomrule
    \end{tabular}}
\end{table}

\begin{table}[tbp]
\centering
    \caption{Comparison of \approach against the Baselines in Terms of \textbf{Execution Pass Rate (EPR)}}
    \label{rq1b}
    \resizebox{\textwidth}{!}{
    \begin{tabular}{lrrrrrrrrrrr}
    \toprule
    \textbf{Project} & \multicolumn{1}{c}{hjson-cpp} & \multicolumn{1}{c}{tinyxml2} & \multicolumn{1}{c}{yaml-cpp} & \multicolumn{1}{c}{re2} & \multicolumn{1}{c}{ninja} & \multicolumn{1}{c}{leveldb} & \multicolumn{1}{c}{json.cpp} & \multicolumn{1}{c}{glomap} & \multicolumn{1}{c}{{papy}} & \multicolumn{1}{c}{{mlx}} & \multicolumn{1}{c}{\textbf{Avg.}} \\
    \midrule
    \textbf{CodeGeeX4} & 0.00\% & 6.00\% & 2.76\% & 0.09\% & 3.82\% & 0.36\% & 8.96\% & 15.54\% & {40.54\%} & {41.89\%} & 12.00\% \\
    \textbf{DeepSeek-V3} & 0.00\% & 0.00\% & 13.78\% & 0.00\% & 6.69\% & 12.02\% & 24.02\% & 39.91\% & {18.75\%} & {66.29\%} & 18.15\% \\
    \textbf{{GPT-3.5}} & 0.00\% & 0.00\% & \underline{18.37\%} & \underline{3.41\%} & 4.04\% & 8.66\% & 30.82\% & 10.48\% & {38.89\%} & {69.50\%} & 18.42\% \\
    \textbf{GPT-4o} & \underline{14.37\%} & 38.46\% & 15.75\% & 0.00\% & 6.57\% & 7.61\% & 49.10\% & \underline{47.88\%} & {\underline{46.22\%}} & {\underline{69.94\%}} & \underline{29.59\%} \\
    \textbf{ChatTester} & 12.00\% & 13.95\% & 4.91\% & 1.75\% & 8.77\% & 0.00\% & 9.38\% & 8.82\% & {17.14\%} & {13.27\%} & 9.00\% \\
    \textbf{HITS} & 1.10\% & 5.45\% & 4.52\% & 0.00\% & \underline{14.90\%} & 1.73\% & 0.00\% & 1.92\% & {0.00\%} & {0.00\%} & 2.96\% \\
    {\textbf{\testpilot}} & {6.67\%} & {\underline{64.55\%}} & {18.14\%} & {0.00\%} & {13.33\%} & {\underline{23.98\%}} & {\underline{70.30\%}} & {12.00\%} & {34.72\%} & {2.66\%} & {24.64\%} \\
    \midrule
    \textbf{\approach} & \textbf{77.50\%} & \textbf{88.24\%} & \textbf{67.82\%} & \textbf{49.37\%} & \textbf{62.12\%} & \textbf{41.00\%} & \textbf{98.91\%} & \textbf{78.02\%} & {\textbf{81.43\%}} & {\textbf{89.13\%}} & \textbf{73.35\%} \\
    \bottomrule
    \end{tabular}}
\end{table}

\begin{table}[tbp]
\centering
    \caption{Comparison of \approach against the Baselines in Terms of \textbf{Line Coverage ($\mathbf{Cov_L}$)}}
    \label{rq1c}
    \resizebox{\textwidth}{!}{
    \begin{tabular}{lrrrrrrrrrrr}
    \toprule
    \textbf{Project} & \multicolumn{1}{c}{hjson-cpp} & \multicolumn{1}{c}{tinyxml2} & \multicolumn{1}{c}{yaml-cpp} & \multicolumn{1}{c}{re2} & \multicolumn{1}{c}{ninja} & \multicolumn{1}{c}{leveldb} & \multicolumn{1}{c}{json.cpp} & \multicolumn{1}{c}{glomap} & \multicolumn{1}{c}{{papy}} & \multicolumn{1}{c}{{mlx}} & \multicolumn{1}{c}{\textbf{Avg.}} \\
    \midrule
    \textbf{CodeGeeX4} & 0.00\% & 15.30\% & 16.54\% & 0.00\% & 8.44\% & 10.45\% & 7.14\% & 21.75\% & {7.22\%} & {8.11\%} & 9.50\% \\
    \textbf{DeepSeek-V3} & 0.00\% & 0.00\% & \underline{31.50\%} & 0.00\% & \underline{20.94\%} & 22.23\% & 37.31\% & \underline{30.04\%} & {\underline{27.78\%}} & {16.80\%} & 18.66\% \\
    \textbf{{GPT-3.5}} & 0.00\% & 0.00\% & 25.59\% & 0.84\% & 10.68\% & 14.73\% & 12.09\% & 16.75\% & {22.10\%} & {17.60\%} & 12.04\% \\
    \textbf{GPT-4o} & \underline{7.73\%} & 43.21\% & 30.17\% & 0.00\% & 10.64\% & 15.45\% & 26.50\% & 22.31\% & {22.89\%} & {\underline{17.85\%}} & \underline{19.68\%} \\
    \textbf{ChatTester} & 1.68\% & 45.00\% & 0.00\% & \underline{3.05\%} & 17.52\% & 0.00\% & 3.56\% & 4.47\% & {9.93\%} & {4.58\%} & 8.98\% \\
    \textbf{HITS} & 0.00\% & 35.64\% & 25.56\% & 0.00\% & 19.55\% & 8.53\% & 0.00\% & 4.15\% & {7.02\%} & {0.00\%} & 9.34\% \\
    {\textbf{\testpilot}} & {0.00\%} & {\underline{70.90\%}} & {0.00\%} & {0.00\%} & {0.00\%} & {\underline{24.33\%}} & {\underline{48.17\%}} & {0.00\%} & {22.11\%} & {4.55\%} & {17.01\%} \\
    \midrule
    \textbf{\approach} & \textbf{27.94\%} & \textbf{77.61\%} & \textbf{44.46\%} & \textbf{42.56\%} & \textbf{56.60\%} & \textbf{25.25\%} & \textbf{50.71\%} & \textbf{42.15\%} & {\textbf{53.11\%}} & {\textbf{24.20\%}} & \textbf{44.46\%} \\
    \bottomrule
    \end{tabular}}
\end{table}

\subsubsection{Experimental Metric Evaluation}
\label{eme}

Table~\ref{rq1a}, Table~\ref{rq1b}, Table~\ref{rq1c}, and Table~\ref{rq1d} present the performance of \approach and selected baselines in C++ unit test generation, evaluated across both correctness metrics (i.e., \textbf{CSR} and \textbf{EPR}) and coverage metrics (i.e., \textbf{$\mathbf{Cov_L}$} and \textbf{$\mathbf{Cov_B}$}). The best result for each metric is highlighted in \textbf{bold}, while the second-best result is \underline{underlined}. Our experiments yield the following key findings:
\begin{enumerate}
    \item \textbf{\approach demonstrates superior performance compared to state-of-the-art baselines on the collected benchmark.} When compared to the seven baselines, \approach achieves the best results across all evaluation metrics. A closer examination of the results reveals that the second-best outcomes for each metric vary depending on the project under test. Nevertheless, GPT-4o consistently outperforms other baselines regarding average scores (listed in the \textbf{Avg.} rows) of all the four evaluation metrics. Specifically, \approach surpasses the best baseline, GPT-4o, by 51.55\% in \textbf{CSR}, 43.76\% in \textbf{ERP}, 24.78\% in \textbf{$\mathbf{Cov_L}$}, and 21.55\% in \textbf{$\mathbf{Cov_B}$}. These improvements underscore the effectiveness of \approach in the C++ unit test generation task.

\begin{table}[tbp]
\centering
    \caption{Comparison of \approach against the Baselines in Terms of \textbf{Branch Coverage ($\mathbf{Cov_B}$)}}
    \label{rq1d}
    \resizebox{\textwidth}{!}{
    \begin{tabular}{lrrrrrrrrrrr}
    \toprule
    \textbf{Project} & \multicolumn{1}{c}{hjson-cpp} & \multicolumn{1}{c}{tinyxml2} & \multicolumn{1}{c}{yaml-cpp} & \multicolumn{1}{c}{re2} & \multicolumn{1}{c}{ninja} & \multicolumn{1}{c}{leveldb} & \multicolumn{1}{c}{json.cpp} & \multicolumn{1}{c}{glomap} & \multicolumn{1}{c}{{papy}} & \multicolumn{1}{c}{{mlx}} & \multicolumn{1}{c}{\textbf{Avg.}} \\
    \midrule
    \textbf{CodeGeeX4} & 0.00\% & 9.97\% & 15.73\% & 0.00\% & 9.18\% & 7.88\% & 10.81\% & 22.22\% & {1.35\%} & {2.70\%} & 7.98\% \\
    \textbf{DeepSeek-V3} & 0.00\% & 0.00\% & \underline{24.24\%} & 0.00\% & \underline{19.03\%} & \underline{23.24\%} & 32.58\% & \underline{28.89\%} & {\underline{15.78\%}} & {13.20\%} & 15.70\% \\
    \textbf{{GPT-3.5}} & 0.00\% & 0.00\% & 21.61\% & 0.51\% & 10.22\% & 12.50\% & 13.06\% & 13.33\% & {8.82\%} & {13.32\%} & 9.34\% \\
    \textbf{GPT-4o} & \underline{8.82\%} & 37.18\% & 23.54\% & 0.00\% & 8.97\% & 14.06\% & 24.84\% & 17.78\% & {12.87\%} & {\underline{13.40\%}} & \underline{16.15\%} \\
    \textbf{ChatTester} & 0.89\% & 38.45\% & 0.00\% & \underline{2.24\%} & 15.46\% & 0.00\% & 1.61\% & 2.48\% & {4.23\%} & {2.27\%} & 6.76\% \\
    \textbf{HITS} & 0.00\% & 26.74\% & 17.70\% & 0.00\% & 16.46\% & 8.72\% & 0.00\% & 3.09\% & {0.00\%} & {0.00\%} & 7.27\% \\
    {\textbf{\testpilot}} & {0.00\%} & {\underline{57.14\%}} & {0.00\%} & {0.00\%} & {0.00\%} & {21.48\%} & {\underline{45.48\%}} & {0.00\%} & {8.83\%} & {2.27\%} & {13.52\%} \\
    \midrule
    \textbf{\approach} & \textbf{24.29\%} & \textbf{65.66\%} & \textbf{35.56\%} & \textbf{35.64\%} & \textbf{49.72\%} & \textbf{24.64\%} & \textbf{46.61\%} & \textbf{37.82\%} & {\textbf{38.97\%}} & {\textbf{18.05\%}} & \textbf{37.70\%} \\
    \bottomrule
    \end{tabular}}
\end{table}

\begin{table}[tbp]
    \centering
    \caption{Comparison of \approach against the Baselines in Terms of the Number of Generated Test Cases}
    \label{rq1e}
    \resizebox{\textwidth}{!}{
    \begin{tabular}{lrrrrrrrrrrrr}
    \toprule
    \textbf{Project} & \multicolumn{1}{c}{hjson-cpp} & \multicolumn{1}{c}{tinyxml2} & \multicolumn{1}{c}{yaml-cpp} & \multicolumn{1}{c}{re2} & \multicolumn{1}{c}{ninja} & \multicolumn{1}{c}{leveldb} & \multicolumn{1}{c}{json.cpp} & \multicolumn{1}{c}{glomap} & \multicolumn{1}{c}{{papy}} & \multicolumn{1}{c}{{mlx}} & \multicolumn{1}{c}{\textbf{Total}} & \multicolumn{1}{c}{{\textbf{Failed}}} \\
    \midrule
    \textbf{CodeGeeX4} & 175 & 417 & 652 & 1133 & 1231 & 1115 & 402 & 193 & {74} & {439} & 5831 & {5456 (93.57\%)} \\
    \textbf{DeepSeek-V3} & 463 & 1405 & 1147 & 1466 & 2273 & 1572 & 533 & 233 & {160} & {482} & 9734 & {8664 (89.01\%)} \\
    \textbf{{GPT-3.5}} & 59 & 172 & 827 & 587 & 1139 & 1235 & 292 & 105 & {90} & {458} & 4964 & {4185 (84.30\%)} \\
    \textbf{GPT-4o} & 167 & 143 & 750 & 777 & 1218 & 986 & 277 & 165 & {119} & {479} & 5081 & {4124 (81.16\%)} \\
    \textbf{ChatTester} & 25 & 172 & 265 & 171 & 308 & 163 & 32 & 34 & {35} & {211} & 1416 & {1306 (92.23\%)} \\
    \textbf{HITS} & 181 & 404 & 1239 & 1190 & 1389 & 752 & 233 & 156 & {78} & {1033} & 6655 & {6352 (95.45\%)} \\
    {\textbf{\testpilot}} & {15} & {299} & {215} & {94} & {225} & {955} & {266} & {25} & {72} & {188} & {2354} & {1642 (69.75\%)} \\
    \midrule
    \textbf{\approach} & 40 & 255 & 289 & 140 & 565 & 922 & 183 & 91 & {70} & {230} & 2785 & {1021 (36.66\%)} \\
    \bottomrule
    \end{tabular}}
\end{table}

    \item \textbf{\approach effectively generates fewer compilation errors than the selected baselines.} As observed from Table~\ref{rq1a}, it is important to that, in many instances, the selected baselines fail to generate any compilable unit test cases for the projects under test using the basic prompt. This issue arises because projects (e.g., \texttt{hjson-cpp}) do not include the \texttt{gtest} testing framework. Consequently, when the LLMs generate test code without accounting for these configuration dependencies, they default to using the \texttt{gtest} framework, resulting in compilation errors for all generated test cases. {If configuration-dependent information is incorporated into other baseline approaches, a portion of their compilation errors could be resolved and their performance could potentially be improved.} In contrast, the higher \textbf{CSR} achieved by \approach is largely due to the inclusion of additional project dependencies. \approach guides the LLMs to generate syntactically correct test code by directly incorporating relevant configuration dependency information into the prompt, thereby greatly reducing the occurrence of compilation errors.
    \item \textbf{The project's complexity significantly affects the ability of \approach to generate correct test cases with high coverage.} According to the statistical results in the seventh column of Table~\ref{benchmark}, \texttt{tinyxml2} exhibits a smaller proportion of complex methods. Consequently, the test cases generated by \approach for \texttt{tinyxml2} achieve both high syntactical correctness, with a \textbf{CSR} exceeding 95\%, and successful execution, with an \textbf{EPR} above 85\%. This leads to strong coverage metrics, with both \textbf{$\mathbf{Cov_L}$} and \textbf{$\mathbf{Cov_B}$} exceeding 65\%. Conversely, for the more complex project \texttt{re2}, which contains the most complex methods, the correctness of the generated test code is comparatively low, and the coverage metrics are also less satisfactory. Additionally, although \texttt{leveldb} has relatively few complex methods, \approach performs poorly in generating test cases for this project. This underperformance can be attributed to several factors. We will discuss the intricate cases from \texttt{leveldb} in Section~\ref{bcb}.
    \item \textbf{\approach achieves high code coverage with fewer test cases than the baselines.} Table~\ref{rq1e} reports the number of test cases generated by \approach and the baselines for each project. The \textbf{Total} column represents the overall number of test cases generated across all projects. The statistical results indicate that \approach generates fewer test cases compared to the selected baselines (except \textbf{ChatTester} and \textbf{\testpilot}), proving that \approach does not rely on sampling a large number of test cases to achieve advantages. The reason \textbf{ChatTester} generates the fewest test cases lies in its prompt design, which explicitly instructs the LLM to ``\textit{write one test case for each focal method}''. This constraint significantly limits the number of test cases generated by \textbf{ChatTester}, resulting in suboptimal performance of \textbf{ChatTester} across many projects. {Furthermore, we examine the number of failed test cases generated by \approach and the baselines, as reported in the \textbf{Failed} column of Table~\ref{rq1e}. The results show that \approach produces significantly fewer failed test cases compared to the baselines, with only 36.66\% of its generated test cases failing to compile or execute. This finding suggests that \approach is more effective in generating syntactically and semantically correct test cases while maintaining high code coverage.}
    \item \textbf{Comparison of the code complexity and mock usage between \approach and the baselines.} As shown in Table~\ref{rq1f}, the rows \textbf{Avg. Size} and \textbf{Avg. CC} represent the average number of source code lines and the average cyclomatic complexity of the generated test cases, respectively. Overall, the test cases generated by \approach exhibit slightly greater scale and complexity compared to the baselines. This is expected, as generating correct test assertions necessitates a certain level of complex code logic. {In addition, we examine the frequency of mock usage, defined as the proportion of test files that employ mock objects. \approach demonstrates a significantly higher mock usage frequency of 41.73\%, indicating its greater capability in producing test cases that involve mocking. This is particularly important for effectively testing complex methods and achieving comprehensive code coverage. Beyond the quantitative analysis, we assess the adherence of the generated mocks to established best practices \cite{spadini2017mock}. Figure~\ref{mock} presents a \approach-generated test case for \texttt{Plan::AddSubTarget} from the \texttt{ninja} project. In this example, the core logic of the \texttt{Plan} module is retained, while its dependency nodes and edges are replaced with mocks to simulate various conditions such as dirty or ready states. This strategy enables precise and isolated testing without compromising behavioral integrity or introducing excessive mocking.}

\begin{table}[tbp]
    \centering
    \caption{Comparison of \approach against the Baselines in Terms of the Complexity and Mock Frequency of Generated Test Cases}
    \label{rq1f}
    \resizebox{\textwidth}{!}{
    \begin{tabular}{lcccccccc}
        \toprule
        \textbf{Approach} & \textbf{CodeGeeX4} & \textbf{DeepSeek-V3} & \textbf{{GPT-3.5}} & \textbf{GPT-4o} & \textbf{ChatTester} & \textbf{HITS} & {\textbf{\testpilot}} & \textbf{\approach} \\
        \midrule
        \textbf{Avg. Size} & 7.7 & 7.4 & 6.8 & 8.0 & 7.3 & 5.9 & {7.1} & 9.5 \\
        \textbf{Avg. CC} & 1.3 & 1.1 & 1.1 & 1.2 & 1.3 & 1.0 & {1.4} & 1.5 \\
        {\textbf{Mock Frequency}} & {26.93\%} & {25.38\%} & {5.60\%} & {25.15\%} & {15.23\%} & {21.44\%} & {4.19\%} & {41.73\%} \\
        \bottomrule
    \end{tabular}}
\end{table}

\begin{figure}[tbp]
  \centering
  \includegraphics[width=\textwidth]{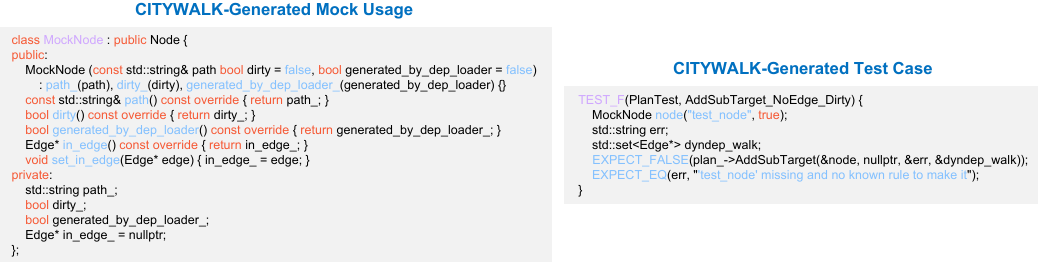}
  \caption{Illustrative Test Case with Compliant Mock Usage.}
  \Description{}
  \label{mock}
\end{figure}

    \item \textbf{The correctness and coverage metrics do not necessarily indicate a positive correlation.} As illustrated in Table~\ref{rq1a}, Table~\ref{rq1b}, Table~\ref{rq1c}, and Table~\ref{rq1d}, high correctness does not always guarantee high code coverage, and low correctness can sometimes lead to higher coverage. For instance, \textbf{DeepSeek-V3} on \texttt{yaml-cpp} demonstrates this phenomenon. Despite achieving lower \textbf{CSR} and \textbf{EPR} compared to other LLM baselines, \textbf{DeepSeek-V3} attains the second-highest code coverage scores. According to the fourth column in Table~\ref{rq1e}, it is evident that \textbf{DeepSeek-V3} generates more test cases for \texttt{yaml-cpp} than other LLM baselines. Consequently, the proportion of incorrect test cases is also relatively high, leading to a scenario characterized by low correctness but high coverage.
    \item \textbf{The two LLM-based unit test generation approaches for Java (i.e., \textbf{ChatTester} and \textbf{HITS}) achieve low performance on the collected C++ projects.} First, the output of LLMs can vary significantly depending on the prompt design. If the prompt does not account for the specific characteristics of different programming languages, the performance of existing Java-specific baselines on C++ unit test generation may not generalize effectively. Furthermore, to ensure fairness in our experiments, all baselines that use LLMs for post-processing perform only a single round of iterative fixes, similar to \approach, which is much fewer than the number of iterations used in their original papers. As a result, the reproduction performance may be lower than the corresponding results reported. However, this also suggests that \approach is less dependent on iterative fixes to achieve its performance.
\end{enumerate}

\begin{figure}[tbp]
  \centering
  \includegraphics[width=\textwidth]{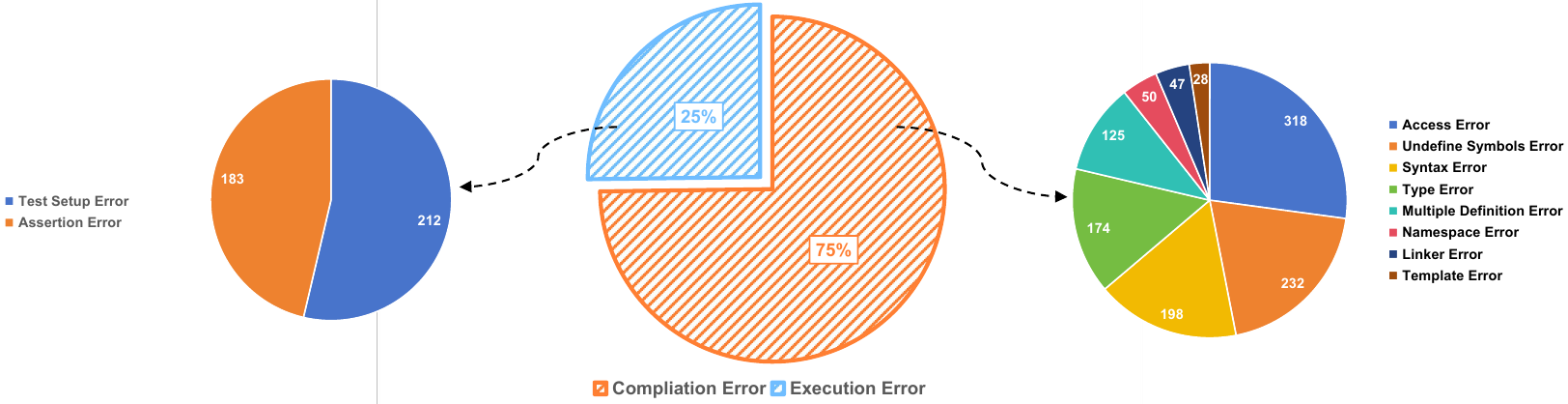}
  \caption{Error Category and Frequency.}
  \Description{}
  \label{rq1g}
\end{figure}

\subsubsection{Bad Case Breakdown}
\label{bcb}

We further conduct an in-depth investigation into the common error categories of the failed test cases generated by \approach. Specifically, we manually analyze the output error messages associated with these failed test cases and summarize the corresponding error categories. {In total, \approach produces 1567 errors across 1021 failed test cases in all ten projects. Since a single test case can contain multiple errors, the total error count exceeds the number of failed test cases.} Our analysis reveals the following findings based on the statistical results:
\begin{enumerate}
    \item \textbf{The two most prevalent categories of errors in the failed test cases are compilation errors and execution errors.} As illustrated in Figure~\ref{rq1g}, we subdivide the two high-level categories into distinct sub-categories. The most common sub-category of compilation errors is \textbf{Access Error}, which occurs when LLMs frequently generate test code that attempts to invalidly access private variables or methods. The most common sub-category of execution errors is \textbf{Test Setup Error}, primarily due to LLMs facing challenges in accurately inferring the mock object configurations, as well as generating valid test data inputs.
    \item \textbf{The complexity of C++ language features poses significant challenges for \approach in generating accurate test cases.} In the design of \approach, we integrate language-specific domain knowledge derived from empirical observations to address common errors, thereby enhancing the performance of the LLM to some extent. However, since the benchmark is collected from a diverse range of real-world projects, project-specific issues may still impede the generation of correct and high-coverage test code. For instance, many functionalities in the large basic software system \texttt{leveldb} rely heavily on external interactions, such as disk I/O operations and file content checks. This dependence often compels LLMs to redefine existing methods in dependent classes or to invoke undefined methods when constructing I/O streams, leading to errors such as \textbf{Multiple Definition Error} and \textbf{Undefined Symbol Error}, which result in low evaluation metrics. Additionally, \texttt{leveldb} involves advanced features like synchronization locks, which significantly increase the complexity and challenge the test generation process. Moreover, the absence of an automated and strict exception handling mechanism, akin to that in Java, complicates the LLM's ability to generate correct test code with exception handling logic. The most frequently occurring assertion errors can be traced back to the lack of supplementary post-processing techniques (e.g., post-hoc fix), which could help prompt LLMs to rectify incorrect assertions in the generated test cases. We are concerned that providing LLMs with error messages that include expected outputs may encourage them to replicate those outputs directly, thus manipulating the compiler into successful execution of the generated test cases. We will discuss these \textit{false positive} cases in detail in Section~\ref{fp}.
\end{enumerate}

\begin{table}[tbp]
    \centering
    \caption{Generalization Results of \approach on Different LLMs in Terms of \textbf{Compilation Success Rate (CSR)}}
    \label{rq1h}
    \resizebox{\textwidth}{!}{
    \begin{tabular}{lrrrrrrrrrrr}
        \toprule
        \multicolumn{1}{c}{\textbf{Project}} & \multicolumn{1}{c}{hjson-cpp} & \multicolumn{1}{c}{tinyxml2} & \multicolumn{1}{c}{yaml-cpp} & \multicolumn{1}{c}{re2} & \multicolumn{1}{c}{ninja} & \multicolumn{1}{c}{leveldb} & \multicolumn{1}{c}{json.cpp} & \multicolumn{1}{c}{glomap} & \multicolumn{1}{c}{{papy}} & \multicolumn{1}{c}{{mlx}} & \multicolumn{1}{c}{\textbf{Avg. $\uparrow$}} \\
        \midrule
        \textbf{CodeGeeX4} & 0.00\% & 6.95\% & 38.98\% & 0.09\% & 4.22\% & 11.75\% & 10.70\% & 35.75\% & {50.00\%} & {51.35\%} & \multirow{2}{*}{\textbf{19.76\%}} \\
        \textbf{CodeGeeX4 w/ \approach} & 43.59\% & 42.11\% & 54.64\% & 4.30\% & 31.45\% & 29.12\% & 36.45\% & 42.44\% & {61.11\%} & {62.22\%} & \\
        \midrule
        \textbf{DeepSeek-V3} & 0.00\% & 0.00\% & 17.79\% & 0.00\% & 7.35\% & 13.55\% & 25.14\% & 51.93\% & {18.75\%} & {68.34\%} & \multirow{2}{*}{\textbf{37.94\%}} \\
        \textbf{DeepSeek-V3 w/ \approach} & 74.62\% & 72.67\% & 43.45\% & 47.83\% & 41.56\% & 42.53\% & 53.89\% & 62.34\% & {70.11\%} & {73.24\%} & \\
        \midrule
        \textbf{{GPT-3.5}} & {0.00\%} & {0.00\%} & {19.34\%} & {5.45\%} & {5.00\%} & {9.64\%} & {31.16\%} & {28.57\%} & {52.22\%} & {72.80\%} & \multirow{2}{*}{\textbf{23.66\%}} \\
        \textbf{{GPT-3.5} w/ \approach} & {31.78\%} & {72.96\%} & {33.69\%} & {20.05\%} & {27.53\%} & {32.45\%} & {45.23\%} & {37.34\%} & {86.67\%} & {73.07\%} & \\
        \bottomrule
    \end{tabular}}
\end{table}

\begin{table}[tbp]
    \centering
    \caption{Generalization Results of \approach on Different LLMs in Terms of \textbf{Execution Pass Rate (EPR)}}
    \label{rq1i}
    \resizebox{\textwidth}{!}{
    \begin{tabular}{lrrrrrrrrrrr}
        \toprule
        \multicolumn{1}{c}{\textbf{Project}} & \multicolumn{1}{c}{hjson-cpp} & \multicolumn{1}{c}{tinyxml2} & \multicolumn{1}{c}{yaml-cpp} & \multicolumn{1}{c}{re2} & \multicolumn{1}{c}{ninja} & \multicolumn{1}{c}{leveldb} & \multicolumn{1}{c}{json.cpp} & \multicolumn{1}{c}{glomap} & \multicolumn{1}{c}{{papy}} & \multicolumn{1}{c}{{mlx}} & \multicolumn{1}{c}{\textbf{Avg. $\uparrow$}} \\
        \midrule
        \textbf{CodeGeeX4} & 0.00\% & 6.00\% & 2.76\% & 0.09\% & 3.82\% & 0.36\% & 8.96\% & 15.54\% & {40.54\%} & {41.89\%} & \multirow{2}{*}{\textbf{21.36\%}} \\
        \textbf{CodeGeeX4 w/ \approach} & 30.77\% & 37.92\% & 42.92\% & 2.15\% & 26.78\% & 25.56\% & 21.78\% & 32.34\% & {55.56\%} & {57.78\%} & \\
        \midrule
        \textbf{DeepSeek-V3} & 0.00\% & 0.00\% & 13.78\% & 0.00\% & 6.69\% & 12.02\% & 24.02\% & 39.91\% & {18.75\%} & {66.29\%} & \multirow{2}{*}{\textbf{33.43\%}} \\
        \textbf{DeepSeek-V3 w/ \approach} & 67.69\% & 64.92\% & 31.82\% & 42.61\% & 36.78\% & 37.63\% & 47.12\% & 51.56\% & {65.52\%} & {70.12\%} & \\
        \midrule
        \textbf{{GPT-3.5}} & {0.00\%} & {0.00\%} & {18.73\%} & {3.41\%} & {4.04\%} & {8.66\%} & {30.82\%} & {10.48\%} & {38.89\%} & {69.50\%} & \multirow{2}{*}{\textbf{23.71\%}} \\
        \textbf{{GPT-3.5} w/ \approach} & {31.78\%} & {70.24\%} & {28.75\%} & {12.98\%} & {32.89\%} & {28.72\%} & {38.47\%} & {27.45\%} & {80.00\%} & {69.94\%} & \\
        \bottomrule
    \end{tabular}}
\end{table}

\begin{table}[tbp]
    \centering
    \caption{Generalization Results of \approach on Different LLMs in Terms of \textbf{Line Coverage ($\mathbf{Cov_L}$)}}
    \label{rq1j}
    \resizebox{\textwidth}{!}{
    \begin{tabular}{lrrrrrrrrrrr}
        \toprule
        \multicolumn{1}{c}{\textbf{Project}} & \multicolumn{1}{c}{hjson-cpp} & \multicolumn{1}{c}{tinyxml2} & \multicolumn{1}{c}{yaml-cpp} & \multicolumn{1}{c}{re2} & \multicolumn{1}{c}{ninja} & \multicolumn{1}{c}{leveldb} & \multicolumn{1}{c}{json.cpp} & \multicolumn{1}{c}{glomap} & \multicolumn{1}{c}{{papy}} & \multicolumn{1}{c}{{mlx}} & \multicolumn{1}{c}{\textbf{Avg. $\uparrow$}} \\
        \midrule
        \textbf{CodeGeeX4} & 0.00\% & 15.30\% & 16.54\% & 0.00\% & 8.44\% & 10.45\% & 7.14\% & 21.75\% & {7.22\%} & {8.11\%} & \multirow{2}{*}{\textbf{14.42\%}} \\
        \textbf{CodeGeeX4 w/ \approach} & 15.78\% & 46.15\% & 35.28\% & 28.04\% & 19.12\% & 12.45\% & 26.89\% & 37.89\% & {8.11\%} & {9.44\%} & \\
        \midrule
        \textbf{DeepSeek-V3} & 0.00\% & 0.00\% & 31.50\% & 0.00\% & 20.94\% & 22.23\% & 37.31\% & 30.04\% & {27.78\%} & {16.80\%} & \multirow{2}{*}{\textbf{21.79\%}} \\
        \textbf{DeepSeek-V3 w/ \approach} & 22.44\% & 78.17\% & 42.80\% & 39.71\% & 52.23\% & 26.63\% & 48.35\% & 43.45\% & {32.87\%} & {17.85\%} & \\
        \midrule
        \textbf{{GPT-3.5}} & {0.00\%} & {0.00\%} & {25.59\%} & {0.84\%} & {10.68\%} & {14.73\%} & {12.09\%} & {16.75\%} & {22.10\%} & {17.60\%} & \multirow{2}{*}{\textbf{21.67\%}} \\
        \textbf{{GPT-3.5} w/ \approach} & {10.01\%} & {74.87\%} & {42.59\%} & {30.62\%} & {32.45\%} & {20.12\%} & {23.56\%} & {32.56\%} & {52.43\%} & {17.85\%} & \\
        \bottomrule
    \end{tabular}}
\end{table}

\begin{table}[tbp]
    \centering
    \caption{Generalization Results of \approach on Different LLMs in Terms of \textbf{Branch Coverage ($\mathbf{Cov_B}$)}}
    \label{rq1k}
    \resizebox{\textwidth}{!}{
    \begin{tabular}{lrrrrrrrrrrr}
        \toprule
        \multicolumn{1}{c}{\textbf{Project}} & \multicolumn{1}{c}{hjson-cpp} & \multicolumn{1}{c}{tinyxml2} & \multicolumn{1}{c}{yaml-cpp} & \multicolumn{1}{c}{re2} & \multicolumn{1}{c}{ninja} & \multicolumn{1}{c}{leveldb} & \multicolumn{1}{c}{json.cpp} & \multicolumn{1}{c}{glomap} & \multicolumn{1}{c}{{papy}} & \multicolumn{1}{c}{{mlx}} & \multicolumn{1}{c}{\textbf{Avg. $\uparrow$}} \\
        \midrule
        \textbf{CodeGeeX4} & 0.00\% & 9.97\% & 15.73\% & 0.00\% & 9.18\% & 7.88\% & 10.81\% & 22.22\% & {1.35\%} & {2.70\%} & \multirow{2}{*}{\textbf{10.93\%}} \\
        \textbf{CodeGeeX4 w/ \approach} & 5.18\% & 33.07\% & 31.23\% & 21.09\% & 21.45\% & 10.72\% & 21.53\% & 32.98\% & {5.25\%} & {6.67\%} & \\
        \midrule
        \textbf{DeepSeek-V3} & 0.00\% & 0.00\% & 24.24\% & 0.00\% & 19.03\% & 23.24\% & 32.58\% & 28.89\% & {15.78\%} & {13.20\%} & \multirow{2}{*}{\textbf{19.30\%}} \\
        \textbf{DeepSeek-V3 w/ \approach} & 23.65\% & 65.98\% & 35.06\% & 31.94\% & 51.89\% & 25.15\% & 45.41\% & 38.12\% & {18.88\%} & {13.87\%} & \\
        \midrule
        \textbf{{GPT-3.5}} & {0.00\%} & {0.00\%} & {21.61\%} & {0.51\%} & {10.22\%} & {12.50\%} & {13.06\%} & {13.33\%} & {8.82\%} & {13.32\%} & \multirow{2}{*}{\textbf{16.86\%}} \\
        \textbf{{GPT-3.5} w/ \approach} & {12.70\%} & {61.39\%} & {34.99\%} & {24.66\%} & {27.12\%} & {17.33\%} & {19.84\%} & {27.46\%} & {23.09\%} & {13.40\%} & \\
        \bottomrule
    \end{tabular}}
\end{table}

\subsubsection{Generalizability Evaluation}

{We further employ \approach to three LLMs, including CodeGeeX4, DeepSeek-V3, and {GPT-3.5}.} Specifically, we conduct ablation experiments on each LLM individually to investigate the impact of querying the corresponding LLM using the prompting strategy and additional contextual guidance designed by \approach. Table~\ref{rq1h}, Table~\ref{rq1i}, Table~\ref{rq1j}, and Table~\ref{rq1k} present the comparison results using the correctness and coverage metrics, respectively. Each LLM's results are displayed in two lines: the first line shows the results when the LLM directly utilizes the basic prompt to generate unit test cases for the given focal methods, and the second line presents the results when integrated with \approach. We derive two key insights from the statistical results: \textbf{\ding{172}} \textbf{Integrating \approach as a complementary plug-in enhances the performance of C++ unit test generation.} As observed, the three evaluated LLMs demonstrate consistent improvements in all four metrics across the ten projects. \textbf{\ding{173}} \textbf{Performance gains scale markedly with parameter count.} For example, the average performance gain of each metric (listed in the \textbf{Avg. $\uparrow$} column) with DeepSeek-V3-671B is substantially greater than that of CodeGeeX4-9B.

\begin{tcolorbox}[size=title]
    \textbf{Answer to RQ1}: In conclusion, \approach exhibits a marked superiority over the LLM-based baselines across all evaluation metrics, highlighting its effectiveness in C++ unit test generation. Furthermore, \approach effectively harnesses the latent intelligence of the LLMs, demonstrating the potential for seamless integration with additional LLMs in a plug-and-play manner.
\end{tcolorbox}

\subsection{Answering RQ2}

To answer this question, we conduct a series of ablation experiments to assess the impact of different designed components within \approach. To ensure the fairness of comparisons, the parameter configurations align with those described in Section~\ref{imp}.

\subsubsection{Ablation Study}

The components within the \approach design include configuration dependencies \({\mathtt{Dep_{c}}}\), cross-file data dependencies \(\mathtt{Dep_{d}}\), intention contexts \(\mathtt{Context_{intent}}\), guidelines of language-specific domain knowledge \(\mathtt{Guideline_{DK}}\), step-by-step instructions \(\mathtt{PROMPT_{step}}\), and post-processing techniques \(\mathtt{Fix_{rule + prompt}}\). Specifically, we perform ablation experiments by removing one component at a time and analyze the performance contribution of each component regarding the coverage metrics $\mathbf{Cov_L}$ and $\mathbf{Cov_B}$. Table~\ref{rq2a} and Table~\ref{rq2b} show the performance contribution results (denoted as the performance degradation values). The greater the coverage scores decline, the larger the contribution of that component. The \textbf{Avg. $\downarrow$} columns present the average results across all projects. When compared to \approach, each ablation model exhibits varying degrees of decline in terms of $\mathbf{Cov_L}$ and $\mathbf{Cov_B}$, indicating that each designed component contributes to the overall enhancement in generating high-quality unit test cases. According to the statistical results presented in Table~\ref{rq2a} and Table~\ref{rq2b}, the top three components that contribute the most to \approach are \({\mathtt{Dep_{c}}}\), \({\mathtt{Dep_{d}}}\), and \(\mathtt{Context_{intent}}\).

As discussed in Section~\ref{eme}, \({\mathtt{Dep_{c}}}\) plays a crucial role in guiding LLMs to generate syntactically correct test code by directly incorporating configuration dependencies. Thus, projects not utilizing \texttt{gtest} (e.g., \texttt{tinyxml2} and \texttt{papy}) would benefit from \({\mathtt{Dep_{c}}}\) as it helps LLMs recognize the absence of this testing framework, thereby preventing hallucinate invocations of frameworks or libraries absent from the project's dependencies. It is worth noting that post-processing techniques (e.g., Pynguin's method injection \cite{lukasczyk2022pynguin}) could also address certain types of compilation errors. In the design of \approach, we integrate both proactive prevention strategies (e.g., explicitly incorporating \({\mathtt{Dep_{c}}}\)) and reactive post-hoc fixes (i.e., \(\mathtt{Fix_{rule + prompt}}\)). Our ablation study reveals that using only post-processing (\textbf{w/o \({\mathtt{Dep_{c}}}\)}) leads to a 29.49\% drop in average $\mathbf{Cov_L}$ and a 24.58\% drop in average $\mathbf{Cov_B}$. In contrast, using only proactive prevention (\textbf{w/o \(\mathtt{Fix_{rule + prompt}}\)}) results in a smaller decrease: 7.03\% in average $\mathbf{Cov_L}$ and 8.60\% in average $\mathbf{Cov_B}$. These results indicate that relying solely on post-hoc fixes is less effective than employing proactive prevention alone.

\begin{figure}[tbp]
  \centering
  \includegraphics[width=\textwidth]{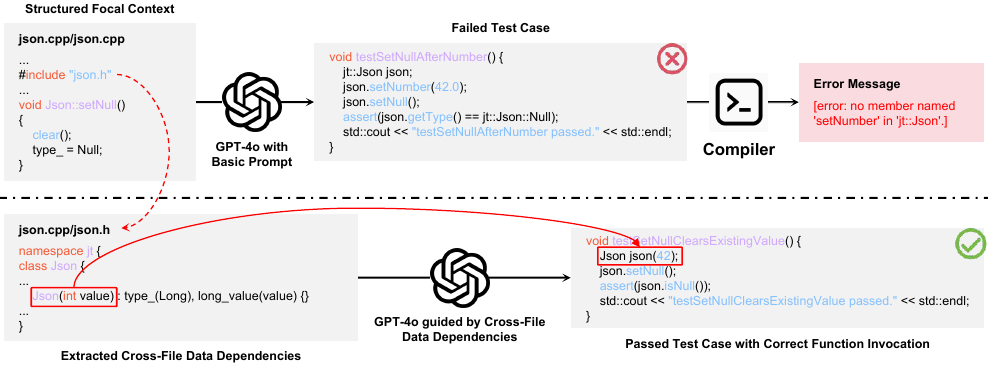}
  \caption{Illustration Example of How Cross-File Data Dependencies Guide LLMs Generate Correct Test Cases.}
  \Description{}
  \label{rq2c}
\end{figure}

\begin{figure}[tbp]
  \centering
  \includegraphics[width=\textwidth]{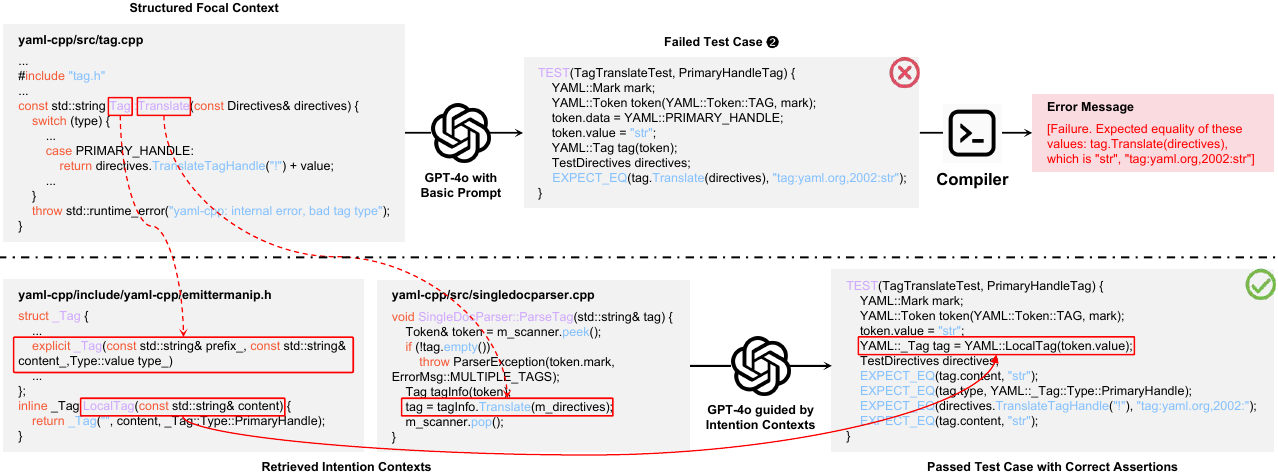}
  \caption{Illustration Example of How Intention Contexts Guide LLMs Generate Correct Test Cases.}
  \Description{}
  \label{rq2d}
\end{figure}

Figure~\ref{rq2c} illustrates how extracted cross-file data dependencies are used as guidance for LLMs to generate correct test cases. By providing the key \(\mathtt{Dep_{d}}\) (i.e., the initialization constructor of \texttt{Json}) from the \texttt{json.h} file, LLMs can prevent compilation errors that would otherwise arise due to invoking the non-existent function \texttt{setNumber}. Figure~\ref{rq2d} illustrates how \(\mathtt{Context_{intent}}\) can guide LLMs in generating correct test cases for the \textbf{Failed Test Case \ding{203}} within Figure~\ref{limitations}. Specifically, the retrieved \texttt{ParseTag} method within the \texttt{singledocparser} class provides an invocation example of the focal method, while the retrieved \texttt{\_Tag} method within the \texttt{emittermanip} header file offers an initialization example of the focal method. These examples significantly aid LLMs in understanding the usage patterns of the focal method, contributing to the generation of functionally correct assertions and enhancing code coverage.

\begin{table}[tbp]
    \centering
    \caption{Performance Contribution of Each Component in Terms of \textbf{Line Coverage ($\mathbf{Cov_L}$)}}
    \label{rq2a}
    \resizebox{\textwidth}{!}{
    \begin{tabular}{lrrrrrrrrrrr}
        \toprule
        \multicolumn{1}{c}{\textbf{Project}} & \multicolumn{1}{c}{hjson-cpp} & \multicolumn{1}{c}{tinyxml2} & \multicolumn{1}{c}{yaml-cpp} & \multicolumn{1}{c}{re2} & \multicolumn{1}{c}{ninja} & \multicolumn{1}{c}{leveldb} & \multicolumn{1}{c}{json.cpp} & \multicolumn{1}{c}{glomap} & \multicolumn{1}{c}{papy} & \multicolumn{1}{c}{mlx} & \multicolumn{1}{c}{\textbf{Avg. $\downarrow$}} \\
        \midrule
        \textbf{\approach} & 27.94\% & 77.61\% & 44.46\% & 42.56\% & 56.60\% & 25.25\% & 50.71\% & 42.15\% & 53.11\% & 24.20\% & \\
        \midrule
        \textbf{\quad w/o \({\mathtt{Dep_{c}}}\)} & -27.94\% & -77.31\% & -12.08\% & -42.56\% & -3.82\% & -1.36\% & -50.71\% & -1.81\% & -53.11\% & -24.20\% & \textbf{-29.49\%} \\
        \textbf{\quad w/o \({\mathtt{Dep_{d}}}\)} & -0.82\% & -32.42\% & -14.50\% & -7.07\% & -15.26\% & -8.69\% & -19.04\% & -13.03\% & -15.51\% & -3.48\% & \textbf{-12.98\%} \\
        \textbf{\quad w/o \({\mathtt{Context_{intent}}}\)} & -12.36\% & -29.84\% & -8.72\% & -8.21\% & -11.21\% & -5.11\% & -6.59\% & -9.70\% & -11.38\% & -0.90\% & \textbf{-10.40\%} \\
        \textbf{\quad w/o \({\mathtt{PROMPT_{step}}}\)} & -5.46\% & -12.97\% & -12.33\% & -9.86\% & -8.42\% & -8.91\% & -12.48\% & -7.59\% & -16.34\% & -7.50\% & \textbf{-10.19\%} \\
        \textbf{\quad w/o \({\mathtt{Guideline_{DK}}}\)} & -16.74\% & -7.40\% & -11.48\% & -7.87\% & -7.04\% & -2.69\% & -3.59\% & -5.59\% & -24.61\% & -6.29\% & \textbf{-9.33\%} \\
        \textbf{\quad w/o \({\mathtt{Fix_{rule + prompt}}}\)} & -5.24\% & -12.51\% & -14.95\% & -8.48\% & -10.76\% & -1.51\% & -1.27\% & -4.26\% & -7.42\% & -3.94\% & \textbf{-7.03\%} \\
        \bottomrule
    \end{tabular}}
\end{table}

\begin{table}[tbp]
    \centering
    \caption{Performance Contribution of Each Component in Terms of \textbf{Branch Coverage ($\mathbf{Cov_B}$)}}
    \label{rq2b}
    \resizebox{\textwidth}{!}{
    \begin{tabular}{lrrrrrrrrrrr}
        \toprule
        \multicolumn{1}{c}{\textbf{Project}} & \multicolumn{1}{c}{hjson-cpp} & \multicolumn{1}{c}{tinyxml2} & \multicolumn{1}{c}{yaml-cpp} & \multicolumn{1}{c}{re2} & \multicolumn{1}{c}{ninja} & \multicolumn{1}{c}{leveldb} & \multicolumn{1}{c}{json.cpp} & \multicolumn{1}{c}{glomap} & \multicolumn{1}{c}{papy} & \multicolumn{1}{c}{mlx} & \multicolumn{1}{c}{\textbf{Avg. $\downarrow$}} \\
        \midrule
        \textbf{\approach} & 24.29\% & 65.66\% & 35.56\% & 35.64\% & 49.72\% & 24.64\% & 46.61\% & 37.82\% & 38.97\% & 18.05\% & \\
        \midrule
        \textbf{\quad w/o \({\mathtt{Dep_{c}}}\)} & -24.29\% & -65.66\% & -10.04\% & -35.64\% & -3.38\% & -1.50\% & -46.61\% & -1.70\% & -38.97\% & -18.05\% & \textbf{-24.58\%} \\
        \textbf{\quad w/o \({\mathtt{Dep_{d}}}\)} & -1.99\% & -30.69\% & -11.48\% & -7.94\% & -14.60\% & -8.75\% & -19.33\% & -13.93\% & -18.61\% & -4.92\% & \textbf{-13.22\%} \\
        \textbf{\quad w/o \({\mathtt{Context_{intent}}}\)} & -14.56\% & -28.48\% & -7.00\% & -9.87\% & -9.60\% & -5.17\% & -6.83\% & -9.59\% & -14.14\% & -2.93\% & \textbf{-10.82\%} \\
        \textbf{\quad w/o \({\mathtt{PROMPT_{step}}}\)} & -5.97\% & -13.60\% & -10.19\% & -9.90\% & -6.16\% & -8.97\% & -12.49\% & -7.59\% & -19.04\% & -9.25\% & \textbf{-10.32\%} \\
        \textbf{\quad w/o \({\mathtt{Guideline_{DK}}}\)} & -15.51\% & -8.07\% & -8.70\% & -9.02\% & -5.83\% & -2.75\% & -3.72\% & -5.70\% & -28.97\% & -4.65\% & \textbf{-9.29\%} \\
        \textbf{\quad w/o \({\mathtt{Fix_{rule + prompt}}}\)} & -6.12\% & -9.68\% & -8.52\% & -9.60\% & -30.33\% & -3.11\% & -0.64\% & -3.26\% & -8.92\% & -5.83\% & \textbf{-8.60\%} \\
        \bottomrule
    \end{tabular}}
\end{table}

\subsubsection{The Impact of Different Error-Fixing Phases within \approach}

We further investigate the influence of individual error-fixing phases within our three-phase post-processing techniques on test case quality enhancement. As described in Section~\ref{postprocessing}, Phase \ding{202} and Phase \ding{203} focus on resolving syntactic and compilation errors via predefined rules, while Phase \ding{204} leverages the LLM to address compilation errors. Through ablation experiments that incrementally incorporate each phase, we quantify their respective contributions using the correctness metrics \textbf{CSR} and \textbf{EPR}. Table~\ref{rq2e} and Table~\ref{rq2f} present phase-by-phase performance improvements, where higher correctness score increments reflect greater phase contributions. The \textbf{Avg. $\uparrow$} columns aggregate cross-project average results. Specifically, the integration of Phase \ding{202} yields improvements of 3.10\% in CSR and 3.19\% in EPR. Subsequent incorporation of Phase \ding{203} delivers substantial enhancements, achieving additional gains of 29.80\% in \textbf{CSR} and 25.98\% in \textbf{EPR}. Including Phase \ding{204} sustains quality improvements with further increases of 8.65\% and 6.75\% in \textbf{CSR} and \textbf{EPR}, respectively. The varying contributions across phases originate from differences in the types of errors resolved. Phase \ding{202} focuses on correcting basic syntactic errors. However, since \approach already incorporates prompt optimization techniques that significantly reduce such errors, this phase yields only marginal improvements. In contrast, Phase \ding{203} utilizes compiler feedback to resolve a broader range of rule-based and compiler-diagnosable errors, resulting in more substantial gains. Finally, Phase \ding{204} targets complex compilation errors that require deeper semantic reasoning about the root causes, thereby further enhancing test case quality. Statistical analysis demonstrates that all three phases exhibit non-negative performance impacts, collectively enhancing test case quality across both metrics.

\begin{table}[tbp]
  \centering
  \caption{{Performance Contribution of Each Error-Fixing Phase in Terms of \textbf{Compilation Success Rate (CSR)}}}
  \label{rq2e}
  \resizebox{\textwidth}{!}{
  \begin{tabular}{lrrrrrrrrrrr}
      \toprule
      \multicolumn{1}{c}{\textbf{{Project}}} & \multicolumn{1}{c}{{hjson-cpp}} & \multicolumn{1}{c}{{tinyxml2}} & \multicolumn{1}{c}{{yaml-cpp}} & \multicolumn{1}{c}{{re2}} & \multicolumn{1}{c}{{ninja}} & \multicolumn{1}{c}{{leveldb}} & \multicolumn{1}{c}{{json.cpp}} & \multicolumn{1}{c}{{glomap}} & \multicolumn{1}{c}{{papy}} & \multicolumn{1}{c}{{mlx}} & \multicolumn{1}{c}{\textbf{{Avg. $\uparrow$}}} \\
      \midrule
      \textbf{{No Error-Fixing Phase}} & {77.08\%} & {67.96\%} & {27.59\%} & {34.76\%} & {22.15\%} & {28.78\%} & {74.56\%} & {36.91\%} & {54.37\%} & {22.97\%} & \\
      \midrule
      \textbf{\quad {w/ Phase \ding{202}}} & {+0.00\%} & {+5.97\%} & {+0.00\%} & {+1.72\%} & {+1.60\%} & {+4.97\%} & {+0.00\%} & {+16.78\%} & {+0.00\%} & {+0.00\%} & {\textbf{+3.10\%}} \\
      \textbf{\quad {w/ Phase \ding{202} \(+\) \ding{203}}} & {+22.92\%} & {+27.97\%} & {+37.32\%} & {+56.04\%} & {+44.52\%} & {+11.39\%} & {+16.58\%} & {+58.33\%} & {+0.00\%} & {+53.90\%} & {\textbf{+32.90\%}} \\
      \textbf{\quad {w/ Phase \ding{202} \(+\) \ding{203} \(+\) \ding{204}}} & {+22.92\%} & {+29.29\%} & {+52.69\%} & {+58.81\%} & {+47.94\%} & {+19.05\%} & {+25.44\%} & {+59.79\%} & {+29.92\%} & {+69.64\%} & {\textbf{+41.55\%}} \\
      \bottomrule
  \end{tabular}}
\end{table}

\begin{table}[tbp]
  \centering
  \caption{{Performance Contribution of Each Error-Fixing Phase in Terms of \textbf{Execution Pass Rate (EPR)}}}
  \label{rq2f}
  \resizebox{\textwidth}{!}{
  \begin{tabular}{lrrrrrrrrrrr}
      \toprule
      \multicolumn{1}{c}{\textbf{{Project}}} & \multicolumn{1}{c}{{hjson-cpp}} & \multicolumn{1}{c}{{tinyxml2}} & \multicolumn{1}{c}{{yaml-cpp}} & \multicolumn{1}{c}{{re2}} & \multicolumn{1}{c}{{ninja}} & \multicolumn{1}{c}{{leveldb}} & \multicolumn{1}{c}{{json.cpp}} & \multicolumn{1}{c}{{glomap}} & \multicolumn{1}{c}{{papy}} & \multicolumn{1}{c}{{mlx}} & \multicolumn{1}{c}{\textbf{{Avg. $\uparrow$}}} \\
      \midrule
      \textbf{{No Error-Fixing Phase}} & {54.17\%} & {61.27\%} & {23.24\%} & {31.33\%} & {19.85\%} & {27.12\%} & {71.93\%} & {27.52\%} & {45.69\%} & {21.70\%} & \\
      \midrule
      \textbf{\quad {w/ Phase \ding{202}}} & {+4.16\%} & {+7.66\%} & {+0.00\%} & {+1.58\%} & {+0.30\%} & {+3.42\%} & {+0.00\%} & {+14.76\%} & {+0.00\%} & {+0.00\%} & {\textbf{+3.19\%}} \\
      \textbf{\quad {w/ Phase \ding{202} \(+\) \ding{203}}} & {+21.51\%} & {+25.61\%} & {+32.46\%} & {+51.43\%} & {+37.27\%} & {+9.44\%} & {+15.05\%} & {+47.48\%} & {+0.00\%} & {+51.43\%} & {\textbf{+29.17\%}} \\
      \textbf{\quad {w/ Phase \ding{202} \(+\) \ding{203} \(+\) \ding{204}}} & {+23.33\%} & {+26.97\%} & {+44.58\%} & {+55.81\%} & {+42.27\%} & {+13.88\%} & {+26.98\%} & {+50.50\%} & {+7.42\%} & {+67.43\%} & {\textbf{+35.92\%}} \\
      \bottomrule
  \end{tabular}}
\end{table}

\begin{tcolorbox}[size=title]
    \textbf{Answer to RQ2}: To sum up, all components of \approach significantly improve the performance of C++ unit test generation in terms of the coverage metrics.
\end{tcolorbox}

\section{Discussion}
\label{dis}

\subsection{Effectiveness of \approach in Bug Detection}

{Detecting real software bugs is a critical criterion for evaluating the effectiveness of automated unit test generation approaches. To complement our assessment based on correctness and coverage metrics, we employ mutation testing to measure the bug-detection capability of \approach-generated test cases. Prior studies \cite{jia2011analysis, fraser2011evosuite} have validated the utility of mutation testing for both evaluating test quality and guiding the generation of more robust test cases. Mutation testing works by introducing small artificial bugs (i.e., \textbf{mutants}) into the program under test. A test suite is considered effective if it can distinguish the mutated version from the original, i.e., ``kill'' the mutant by triggering observable failures. In this study, we adopt the updated open-source tool \textit{universalmutator} \cite{deb2024syntax}, which supports a broad range of mutation operations, as outlined below:}
\begin{itemize}
    \item {\textbf{Arithmetic operator mutations}: e.g., $+$ $\leftrightarrow$ $-$, $\ast$ $\leftrightarrow$ $/$}
    \item {\textbf{Comparison operator mutations}: e.g., $<$ $\leftrightarrow$ $>$, $==$ $\leftrightarrow$ $!=$}
    \item {\textbf{Logical operator mutations}: e.g., \texttt{\&\&} $\leftrightarrow$ \texttt{$||$}}
    \item {\textbf{Control structure mutations}: e.g., removing \texttt{else}, \texttt{break} $\leftrightarrow$ \texttt{continue}}
    \item {\textbf{Structural mutations}: e.g., deleting or commenting out code blocks}
    \item {\textbf{Literal and constant mutations}: e.g., \texttt{true} $\leftrightarrow$ \texttt{false}, \texttt{0} $\leftrightarrow$ \texttt{1}, string substitutions}
\end{itemize}

{To facilitate analysis, we select two single-file projects, \texttt{tinyxml2} and \texttt{json.cpp}, as evaluation subjects. We first perform mutation testing on the test cases generated by \approach, and then calculate the mutation score, which is defined as the ratio of killed mutants to total valid mutants. A higher mutation score indicates stronger bug detection capabilities and higher oracle quality \cite{xie2006augmenting}. We yield the following key observations according to the evaluation results presented in Table~\ref{mutation}:}
\begin{enumerate}
    \item {\approach-generated test cases respective achieve a mutation score of 89.34\% for \texttt{json.cpp} and 81.12\% for \texttt{tinyxml2}. These results indicate that \approach is highly effective in generating test cases that detect a substantial proportion of artificial bugs introduced into the code, validating its capability to uncover potential bugs.}
    \item {The mutation score for \texttt{json.cpp} (created after the GPT-4o knowledge cutoff) outperforms \texttt{tinyxml2} (which has a risk of potential data leakage) by 8.22\%. This provides additional evidence that \approach's ability to generate high-quality test cases is primarily driven by the designed components as guidance, rather than relying on GPT-4o's memorization of training data.}
    \item {According to the results from the SBST 2022 tool competition \cite{schweikl2022evosuite}, EvoSuite achieved a mutation score of 34.1\% on the competition benchmark. Additionally, the open-source tool \textit{universalmutator} achieved an average mutation score of 35\% on the evaluated C++ projects in the corresponding paper \cite{deb2024syntax}. In comparison, the mutation scores achieved by \approach are notably higher, further emphasizing the effectiveness of \approach-generated test cases.}
\end{enumerate}

\begin{table}[tbp]
    \centering
    \caption{{Mutation Testing for \approach-Generated Test Cases on \texttt{tinyxml2} and \texttt{json.cpp}}}
    \label{mutation}
    \begin{tabular}{lrrr}
        \toprule
        \multicolumn{1}{c}{{\textbf{Project}}} & \multicolumn{1}{c}{{\textbf{\# Total Valid Mutants}}} & \multicolumn{1}{c}{{\textbf{\# Killed Mutants}}} & \multicolumn{1}{c}{{\textbf{Mutation Score}}} \\
        \midrule
        {\texttt{tinyxml2}} & {6272} & {5088} & {81.12\%} \\
         {\texttt{json.cpp}} & {844} & {754} & {89.34\%} \\
        \bottomrule
    \end{tabular}
\end{table}

\begin{figure}[tbp]
  \centering
  \includegraphics[width=\textwidth]{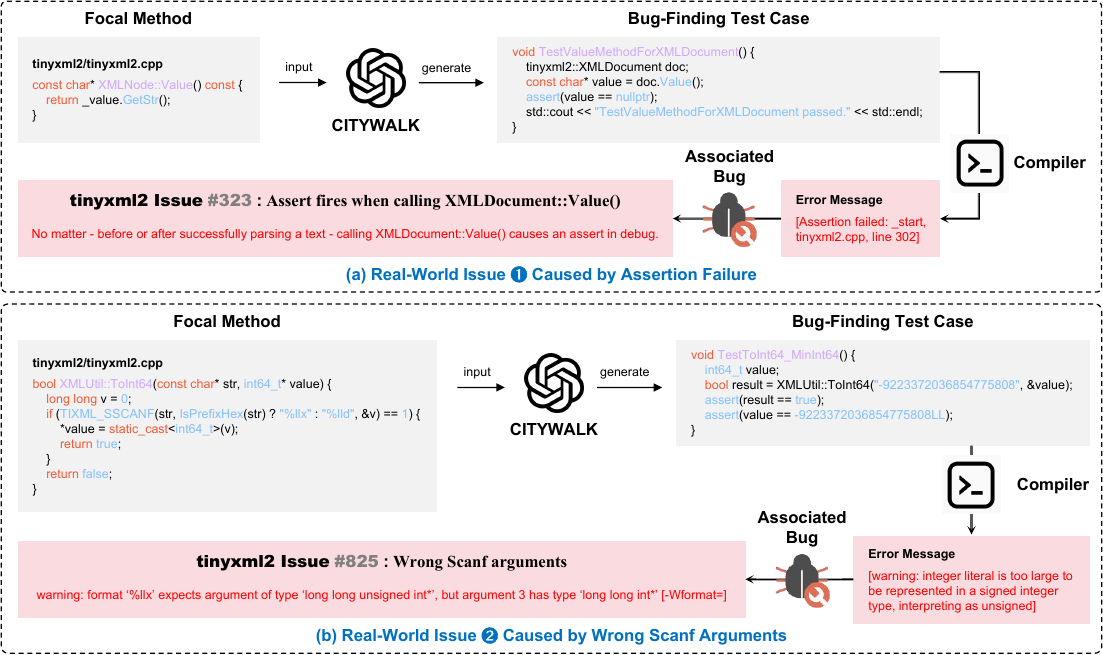}
  \caption{Illustration of Two Real-World Issues from \texttt{tinyxml2} Found by \approach.}
  \Description{}
  \label{issue}
\end{figure}

Furthermore, we conduct a case-by-case analysis to assess the usefulness of \approach in discovering real-world bugs. We manually inspect two bugs from \texttt{tinyxml2} that were identified by \approach-generated test cases. These are real-world issues that have been reported by human developers. As illustrated in Figure~\ref{issue}(a), the \approach-generated test case first creates an \texttt{XMLDocument} object, \texttt{doc}, which then invokes the \texttt{Value} function. This function, in turn, calls \texttt{XMLNode::Value()}. Since the variable \texttt{\_value} within \texttt{XMLNode::Value()} may not have been initialized, it triggers an assertion error in the \texttt{GetStr} function. This bug corresponds to \textbf{Issue \#323}\footnote[15]{https://github.com/leethomason/tinyxml2/issues/323} in \texttt{tinyxml2}, where a human developer also commits an assertion failure when calling \texttt{Value}. Figure~\ref{issue}(b) shows another bug triggered by \approach, where invalid hexadecimal format string arguments are passed to \texttt{ToInt64}. This bug is associated with \textbf{Issue \#825}\footnote[16]{https://github.com/leethomason/tinyxml2/issues/825} in \texttt{tinyxml2}, where a human developer also identifies a similar issue when calling the static member function \texttt{ToInt64}. {In summary, the test cases generated by \approach demonstrate the potential to detect real bugs in open-source projects.}

\subsection{Efficiency of \approach}

{To evaluate the efficiency and practical feasibility of \approach in the context of C++ unit test generation, Table~\ref{efficiency} presents the average execution time per focal method and the corresponding token usage for \approach and each baseline approach. The following insights can be drawn from the statistical results:}
\begin{enumerate}
    \item Despite not being the fastest or most cost-efficient approach, \approach maintains acceptable computational overhead. The average execution time per focal method is 34.59 seconds, significantly lower than the slowest baseline (\textbf{HITS} at 184.39 seconds). {Since writing unit tests is time-consuming \cite{fucci2018longitudinal}, \approach remains a viable option for developers to enhance testing accuracy while ensuring prompt response times.} Moreover, although \approach incurs a higher average token usage (5726 tokens per method) compared to most baselines, {the incremental cost per method is approximately \$0.03 higher than the best baseline \testpilot, which may be justifiable given \approach's improved correctness and coverage—especially in scenarios where test reliability is prioritized (e.g., safety-critical systems).}
    \item {Although \approach does not achieve the lowest values in terms of efficiency metrics, it consistently delivers superior test quality, stemming from the designed contextual guidance and post-processing mechanisms. This indicates that \approach strategically trades a moderate increase in computational and monetary cost for substantial gains in test effectiveness, a trade-off often justified in practical software development scenarios.}
\end{enumerate}

\begin{table}[htbp]
    \centering
    \caption{{Efficiency Comparison of \approach against the Baselines}}
    \label{efficiency}
    \resizebox{\textwidth}{!}{
    \begin{tabular}{lcccccccc}
        \toprule
        {\textbf{Approach}} & {\textbf{CodeGeeX4}} & {\textbf{DeepSeek-V3}} & {\textbf{{GPT-3.5}}} & {\textbf{GPT-4o}} & {\textbf{ChatTester}} & {\textbf{HITS}} & {\textbf{\testpilot}} & {\textbf{\approach}} \\
        \midrule
        {\textbf{Avg. Execution Time (s)}} & {15.93} & {57.02} & {10.26} & {18.62} & {70.30} & {184.39} & {24.45} & {34.59} \\
        {\textbf{Avg. Token Usage}} & {2533} & {3553} & {1849} & {2583} & {2222} & {11736} & {1436} & {5726} \\
        \bottomrule
    \end{tabular}}
\end{table}

\subsection{Readability and Usability of Test Cases Generated by \approach}

{The ultimate goal of automated unit test generation is to assist developers in writing test cases. To compare the readability and usability of the test cases generated by \approach with those produced by four LLM-based baselines (i.e., \textbf{CodeGeeX4}, \textbf{DeepSeek-V3}, \textbf{{GPT-3.5}}, and \textbf{GPT-4o}), we conduct a human evaluation to determine developer preference. We invite five participants, each with over three years of C++ development experience, to perform the assessment. We focus on focal methods for which both \approach and the LLMs generated correct test cases, as recommending test cases that fail to compile or execute is impractical. We further limit our analysis to methods with existing human-written test cases in the original project repositories. Consequently, we randomly select 25 focal methods across five projects from our benchmark. Each participant is required to evaluate 150 test cases, comprising one \approach-generated, four LLM-generated, and one human-written test case for each focal method. Each test case is independently scored across the following four aspects. Scores of each aspect range from 1 (lowest quality) to 3 (highest quality). To avoid bias, participants are not informed about the source of each test case (i.e., whether it is LLM-generated or human-written).}
\begin{itemize}
    \item {\textbf{Naming Intuitiveness.} Clarity and descriptiveness of variable and test method names.}
    \item {\textbf{Code Layout.} Structure, logic, and formatting of the test code.}
    \item {\textbf{Assertion Quality.} Effectiveness and relevance of assertions for validating the focal method.}
    \item {\textbf{Adoption Efforts.} Ease of integrating the test case into real-world usage.}
\end{itemize}

{Table~\ref{readability} summarizes the average scores across all participants. The results show that \approach outperforms all LLM baselines in each of the four aspects. Specifically, \approach surpasses the best baseline by 1.41\% in \textbf{Naming Intuitiveness} (compared to DeepSeek-V3), 10.66\% in \textbf{Code Layout} (compared to GPT-4o), 18.27\% in \textbf{Assertion Quality} (compared to DeepSeek-V3), and 7.14\% in \textbf{Adoption Efforts} (compared to DeepSeek-V3 and GPT-4o). Compared with human-written test cases, participants consider \approach-generated test cases with more intuitive naming and more structured code. However, in terms of usability, \approach-generated test cases demonstrate slightly lower assertion quality and comparable adoption effort. This is expected, as generating high-quality assertions remains an open challenge in LLM-based unit test generation. Overall, the findings highlight the superior readability and practical usability of \approach-generated test cases, underscoring its value as a developer-assistive tool.}

\begin{table}[htbp]
    \centering
    \caption{{Comparison of \approach against the LLM Baselines in Terms of the Readability and Usability of Generated Test Cases}}
    \label{readability}
    \begin{tabular}{lcccc}
        \toprule
        \multirow{2.4}{*}{{\textbf{Approach}}} & \multicolumn{2}{c}{{\textbf{Readability}}} & \multicolumn{2}{c}{{\textbf{Usability}}} \\
        \cmidrule(lr){2-3} \cmidrule(lr){4-5}
        & {\textbf{Naming Intuitiveness}} & {\textbf{Code Layout}} & {\textbf{Assertion Quality}} & {\textbf{Adoption Efforts}} \\
        \midrule
        {\textbf{CodeGeeX4}} & {2.70} & {2.32} & {1.46} & {2.47} \\
        {\textbf{DeepSeek-V3}} & {2.83} & {2.39} & {2.08} & {2.52} \\
        {\textbf{{GPT-3.5}}} & {2.55} & {2.26} & {1.75} & {2.38} \\
        {\textbf{GPT-4o}} & {2.70} & {2.44} & {2.06} & {2.52} \\
        \midrule
        {\textbf{\approach}} & {2.87 (1.41\% \(\uparrow\))} & {2.70 (10.66\% \(\uparrow\))} & {2.46 (18.27\% \(\uparrow\))} & {2.70 (7.14\% \(\uparrow\))} \\
        \midrule
        {\textbf{Human-Written}} & {2.64} & {2.63} & {2.87} & {2.75} \\
        \bottomrule
    \end{tabular}
\end{table}

\subsection{False-Positive Executable Test Cases}
\label{fp}

\begin{figure}[tbp]
  \centering
  \includegraphics[width=\textwidth]{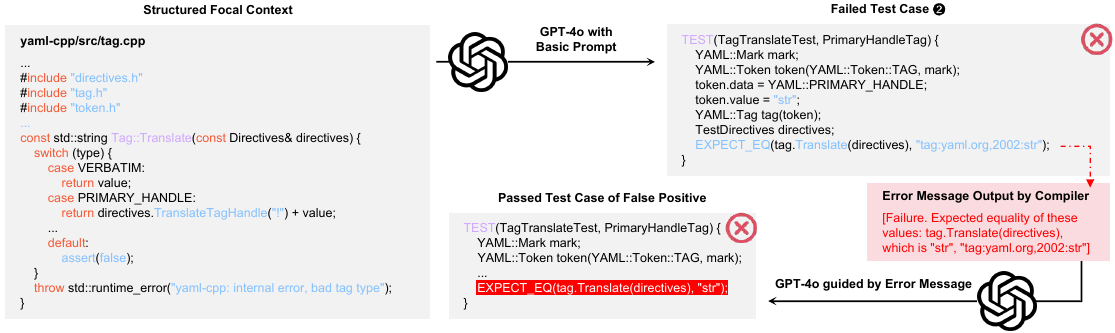}
  \caption{False Positive Example for the Failed Test Case \ding{203} within Figure~\ref{limitations} using Iterative LLM-Based Fixing.}
  \Description{}
  \label{false}
\end{figure}

Existing LLM-based unit test generation approaches \cite{yuan2024evaluating,wang2024hits} primarily rely on compiler error messages to iteratively guide LLMs in fixing test cases that fail during execution. However, this strategy can lead LLMs to align expected values with observed outputs, resulting in test cases that pass without necessarily validating the intended functionality. As illustrated in Figure~\ref{false}, when provided with the error message, GPT-4o directly uses the expected value \texttt{``str''} as the input for assertions, resulting in a passing test case that is, in reality, a false positive. {While such behavior resembles regression testing, a common practice in automated unit test generation, it risks reinforcing incorrect behavior when the underlying implementation contains faults.

Our goal is to highlight this specific risk introduced by compiler message–driven error-fixing strategies in LLM-based frameworks. When LLMs are exposed to runtime or compiler messages that reveal actual outputs, they may inadvertently ``learn'' to generate assertions that merely reproduce observed behavior rather than verify correctness.} Thus, \approach is intentionally designed to avoid relying on compiler feedback for fixing execution failures. Effectively detecting and addressing such false positives remains as our future work.

\subsection{Threats to Validity}
\label{thr}

{In this subsection, we discuss the primary threats to the validity of \approach, as outlined below:}
\begin{itemize}
    \item \textbf{External Threat.} The primary threats to external validity lie in the diversity of the projects used for evaluation and its generalization to other LLMs. In this paper, we collect 1288 focal methods across ten real-world open-source C++ projects crawled from GitHub, ensuring a degree of diversity and quality in our evaluation. For comparison, we select two open-source code LLMs (CodeGeeX4 and DeepSeek-V3) and two closed-source commercial LLMs ({GPT-3.5} and GPT-4o), considering their varied model sizes, architectures, and effectiveness on coding tasks. Future work will involve expanding the benchmark and integrating additional LLMs to better assess the generalizability of \approach.
    \item {\textbf{Internal Threat.} LLMs exhibit sensitivity to prompt configuration and hyper-parameter settings, particularly the number of task examples and the phrasing of natural language instructions, which can substantially influence performance. To ensure a fair comparison, we use consistent prompts and hyper-parameters across both \approach and all baseline approaches. Additionally, we adopt empirically motivated default configurations rather than fine-tuning prompts or parameters through trial-and-error. We acknowledge that further improvements may be achievable through additional prompt and hyper-parameter tuning. Another potential threat to validity relates to data leakage issue. As GPT-4o is a closed-source model, the exact composition of its training data are not publicly disclosed. Despite this limitation, \approach exhibits a significant improvement in generating high-quality test cases compared to GPT-4o, which utilizes the same underlying architecture. These enhancements suggest that the performance gains achieved by \approach are not merely attributable to the model's memorization of its training data.}
\end{itemize}

\section{Related Work}
\label{rel}

To mitigate the manual effort associated with writing unit tests for developers in practice, researchers have proposed various automation techniques aimed at enhancing testing efficiency. Existing approaches can be broadly categorized into the following three technical avenues.

\subsection{Program Analysis-Based Automated Unit Testing Tools}

EvoSuite \cite{fraser2011evosuite} is an automated test case generation tool tailored for Java. It utilizes mutation testing and constraint-solving techniques to generate appropriate assertions, which effectively summarize the behavior of the program while maximizing the number of killed mutants. In contrast, Randoop \cite{pacheco2007feedback,pacheco2007randoop} is an automated tool that adopts feedback-driven random testing techniques to generate assertions. Randoop leverages the outcomes of test executions to generate assertions that accurately capture the program's behavior. Pynguin \cite{lukasczyk2022pynguin}, an extendable test generation framework for Python to produce regression tests via search-based techniques. Coyote C++ \cite{rho2023coyote} employs a sophisticated concolic execution-based method to facilitate fully automated unit testing for C and C++. Despite demonstrating commendable performance in achieving high code coverage, the aforementioned tools exhibit certain limitations: (1) The test code generated by existing tools often suffers from poor readability \cite{daka2015modeling}; (2) The assertions produced by these tools are frequently insufficient in effectively detecting real-world faults \cite{shamshiri2015automatically,shamshiri2015automated}; (3) Search-based techniques may encounter path explosion issues due to excessively large search spaces \cite{tang2024chatgpt}.

\subsection{Pre-Trained Language Model-Based Automated Unit Test Generation}

Tufano et al. \cite{tufano2021unit} pre-trained a language model on large-scale unsupervised Java corpora, subsequently fine-tuning the model for unit test generation, thereby enabling the efficient generation of test cases. Zhang et al. \cite{zhang2023saga} adopted the summarization of focal methods as complementary information to capture the developers' intent, which aids in generating meaningful test assertions for helping developers in writing accurate unit test cases. Similarly, Alagarsamy et al. \cite{alagarsamy2024a3test} utilized both focal methods and assertion declarations during model pre-training, aiming to establish connections between focal methods and corresponding test cases. Shin et al. \cite{shin2024domain} developed project-specific datasets for domain adaptation by leveraging existing developer-written test cases within each project, promoting the generation of more human-like unit tests. Steenhoek et al. \cite{steenhoek2023reinforcement} employed reinforcement learning for model optimization, designing reward functions based on the static quality metrics of the generated unit test cases. Nonetheless, unit test cases generated through the paradigm of pre-training and fine-tuning frequently encounter compilation or execution failures \cite{yang2022survey}.

\subsection{LLM-based Automated Unit Test Generation}

With the emergence of generative artificial intelligence, researchers have increasingly investigated LLM-driven approaches for automatically generating unit test cases \cite{kang2023large,wang2024software}. ChatUniTest \cite{chen2024chatunitest} adaptively constructs dependency contexts for the focal method, and employs a generate-validate-fix mechanism to address errors in the generated test cases. Similarly, ChatTester \cite{yuan2024evaluating} enhances the quality of generated test cases via intent comprehension and iterative correction. {In contrast to prompt engineering-based approaches that merely append file-level dependencies as additional contextual information, \approach conducts a comprehensive analysis of the project under test to extract project-level dependencies that impact test case generation, such as environment requirements. Moreover, \approach employs RAG techniques to integrate language-specific knowledge from project documentation and source code, thereby enhancing performance.} On the other hand, SymPrompt \cite{ryan2024code} guides LLMs to generate high-coverage test code by incorporating symbolic execution-based path information into the prompts. However, SymPrompt does not address the issue of path reachability, which can result in the inclusion of unreachable path information in the prompts, ultimately leading to inaccurate model reasoning. HITS \cite{wang2024hits} simplifies the analysis of complex focal methods through program decomposition and achieves high coverage scores by prompting LLMs to generate test cases for sliced code segments. Unlike previous approaches that primarily target interpreted languages such as Java and Python, \approach is specifically designed to address C++-specific challenges. In this work, \approach conducts an empirical analysis to identify common error patterns in LLM-generated C++ unit tests, It then documents these C++-specific failure patterns as insights to guide the post-processing of generated test cases, ensuring their accuracy and reliability.

\section{Conclusion and Future Work}
\label{con}

This paper presents a novel framework \approach designed to enhance the capabilities of LLMs in generating high-quality C++ unit test cases. We explore the potential of GPT-4o by integrating program analysis techniques with retrieval-augmented strategies, providing guidance through three key aspects: \textbf{project dependencies}, \textbf{intention contexts}, and \textbf{language-specific knowledge}. Additionally, we decompose the unit test generation task into three distinct stages, utilizing step-by-step instructions to streamline the generation of C++ test cases, complemented by effective post-processing techniques. Extensive experiments demonstrate the superiority of \approach, and further ablation studies validate the contributions of each designed component within \approach.

Future work will focus on improving LLM-generated unit test quality through two key advancements: (1) developing robust assertion verification techniques to validate functional correctness beyond coverage metrics, ensuring precise alignment with program specifications, and (2) enhancing bug-detection capabilities to identify diverse real-world software faults. These efforts aim to substantially narrow the gap between LLM-generated and human-written tests in terms of reliability and practical utility.

\begin{acks}
We would like to thank the reviewers for their insightful comments and suggestions. This work was partially supported by the National Key R\&D Program of China (Grant No. 2024YFB4505902), the Major Project of ISCAS (Grant No. ISCAS-ZD-202302), the Basic Research Project of ISCAS (Grant No. ISCAS-JCZD-202403), the Youth Innovation Promotion Association of the Chinese Academy of Sciences (Grant Nos. Y2022044 and 2023121).
\end{acks}

\bibliographystyle{ACM-Reference-Format}
\bibliography{sample-base}

\end{document}